\documentclass{aa}  

\usepackage{graphicx}
\usepackage{txfonts}
\usepackage{arydshln}
\usepackage{lipsum}
\usepackage{ulem}
\usepackage{natbib}    
\usepackage{subcaption}  
\usepackage[colorlinks=true,linkcolor=blue,citecolor=blue,urlcolor=blue]{hyperref}

\usepackage{lscape}             
\usepackage{placeins}           

\let\linenumbers\relax
\let\runningpagewiselinenumbers\relax
\let\columnwiselinenumbersfalse\relax

\begin{document}

   \title{Self-consistent modelling of neutrino production in
   turbulent black hole coronae}

   \author{S. Le Bihan\inst{1}
        \and M. Lemoine\inst{1}
        \and F. Rieger\inst{2,3}}

   \institute{Université Paris Cité, CNRS, Astroparticule et Cosmologie, F-75013 Paris, France\\
             \email{lebihan@apc.in2p3.fr}\\
             \email{mlemoine@apc.in2p3.fr}
            \and
            Max Planck Institute for Plasma Physics (IPP), Boltzmannstraße 2, 85748 Garching, Germany
            \and
            Institute for Theoretical Physics, Heidelberg University, Philosophenweg 12, 69120 Heidelberg, Germany\\
            \email{frank.rieger@ipp.mpg.de} }

   \date{Received XX XX, 20XX}
   \titlerunning{Self-consistent modelling of neutrino production in coronae}

  \abstract
{Stochastic particle acceleration in magnetised turbulent plasmas has emerged as a key mechanism to explain multi-messenger signals from compact astrophysical environments. Self-consistent modelling remains challenging because it requires one to treat several non-linear kinetic processes simultaneously, especially turbulence-driven acceleration and its feedback on the turbulent cascade, as well as the radiative and hadronic losses, including the reprocessing of electromagnetic radiation in radiatively dense environments. The present paper introduces the hybrid numerical code \texttt{Turb-AM3}, which is designed for this purpose. This hybrid numerical code couples the state-of-the-art time-dependent lepto-hadronic radiative solver \texttt{AM3} with a stochastic acceleration module that incorporates recent theoretical advances in turbulent acceleration and accounts for the dynamical damping of turbulence by accelerated particles. In the second part of the paper, we use this code to provide self-consistent time-dependent models of proton acceleration in the turbulent black hole corona of NGC~1068. We find that the IceCube neutrino signal is well reproduced for a standard set of physical parameters describing the black hole corona. The same template model accounts for in a satisfactory way for IceCube observations of other active galactic nuclei. Furthermore, our exploration of parameter space  allowed us to predict detailed template spectral shapes for the TeV neutrino spectrum, which in turn help clarify how future neutrino observations can constrain the properties of turbulent AGN coronae and the underlying acceleration mechanism. This \texttt{Turb-AM3} framework provides a powerful tool to model multi-messenger emission in a broad variety of compact astrophysical environments.}


\maketitle

\section{Introduction}\label{sec:introd}

In 2022, the IceCube Collaboration reported a $4.2\sigma$ statistical excess of high-energy neutrinos in the $\sim 1$-$10$ TeV range spatially associated with the nearby Seyfert 2 galaxy NGC~1068 \citep{aartsenTimeIntegratedNeutrinoSource2020a,icecubecollaborationEvidenceNeutrinoEmission2022}. If confirmed, it will represent a dramatic step forward in our exploration of the high-energy multi-messenger sky. While this active galactic nucleus (AGN) currently provides the most compelling source of high-energy neutrinos, statistical evidence of correlation with other nearby Seyfert galaxies \citep[e.g. NGC~4151, NGC~7469, $\sim3.0 \sigma$ for both and at $E_\nu \simeq 5-50$ TeV and $E_\nu \simeq 50-100$ TeV, respectively;][]{abbasiEvidenceNeutrinoEmission2025} has been increasing in recent years \citep{abbasiSearchNeutrinoEmission2024,abbasiseyfert2025}, suggesting that non-jetted AGN environments are significant contributors to the extragalactic neutrino flux \citep{padovaniHighenergyNeutrinosVicinity2024,neronovNeutrinoSignalPopulation2024,2025ApJ...989..215F,2026arXiv260220145M,2026arXiv260220969Y}. These results expand the landscape of possible sources beyond the flaring blazar TXS 0506+056~\citep{icecubecollaborationNeutrinoEmissionDirection2018,icecubeMultimessengerObservationsFlaring2018}, other jetted candidates such as PKS 1424-240 and GB6 J1542+61\citep{icecubecollaborationEvidenceNeutrinoEmission2022}, and tidal disruption events~\citep{Stein_2021, Reusch_2022,lu2025investigatingcorrelationztftdes}. 

NGC 1068 is a long-studied source due to its relative proximity of $\simeq 10\,$Mpc \citep{tullyExtragalacticDistanceDatabase2009,lianouDustPropertiesStar2019a}. Low-frequency observations of its nucleus point to heavy obscuration below the megaelectronvolt (MeV) range by a dense dusty torus that absorbs the primary radiation from the accretion disk and corona and re-emits it in the infrared band \citep[e.g.][]{jaffeCentralDustyTorus2004}. As a consequence, the intrinsic emission from the innermost regions of the AGN cannot be directly probed at X-ray energies. In the $\gamma-$ray domain, observations with \textit{Fermi}-LAT and MAGIC \citep{thefermicollaborationFermiLargeArea2020,magiccollaborationConstraintsGammarayNeutrino2019} place stringent upper limits on the $\gamma$-ray flux that are at least an order of magnitude below the neutrino flux inferred by IceCube. This striking mismatch between the neutrino and $\gamma$-ray emissions is widely interpreted as evidence that neutrinos are produced in a compact region that is opaque to high-energy photons, due to efficient $\gamma\gamma$ absorption in an intense radiation field~\citep{inoueHighenergyParticlesAccretion2019,  muraseHiddenCoresActive2020, 2021ApJ...922...45K,2022ApJ...941L..17M,eichmannSolvingMultimessengerPuzzle2022,ajelloDisentanglingHadronicComponents2023a,fangHighenergyNeutrinosInner2023,padovaniHighenergyNeutrinosVicinity2024,muraseSubGeVGammaRays2024,dasRevealingProductionMechanism2024}. While alternative explanations have been proposed \citep[e.g.][]{herreraPlausibleIndicationGammaRay2025}, the most compelling interpretation points towards a dense, compact emission zone, most plausibly the turbulent AGN corona of the supermassive black hole (SMBH). The observed multi-messenger signals from NGC 1068 therefore provide a unique indirect probe of particle acceleration to very high energies and radiative processes in the immediate environment of black holes. 

In this emerging context, proton acceleration to $\sim 10-100\,$TeV in AGN cores has received increased attention in recent years, whether around accretion shock waves, in magnetic reconnection layers, in the magnetised turbulence, in shear layers at the jet base, or a combination of these processes (see the references in the previous paragraph as well as e.g. \citealt{fiorilloTeVNeutrinosHard2024,mbarekInterplayAcceleratedProtons2024,lemoineNeutrinosStochasticAcceleration2025,2024arXiv240613336A,2025ApJ...995..166Y,Yuan2026,2025JCAP...04..075K,2026arXiv260101999N,Testagrossa2026}). As AGN coronae are generically regarded as magnetically dominated and highly turbulent environments, stochastic acceleration in the magnetised turbulence emerges as a natural and promising mechanism, and we focus on this scenario in the present paper.

Theoretical models of AGN coronae have traditionally focused on the energisation processes of leptons (electrons and positrons) and their radiative processes, notably inverse Compton scattering and pair production, shaping the X-ray spectra of the AGN \citep[e.g.][for a few recent references]{2015MNRAS.451.4375F,2017MNRAS.467.2566F,beloborodovRadiativeMagneticReconnection2017,2024FrASS..1008056K}. The high compactness of AGN coronae implies that electrons cool on short timescales $\ll R_{\rm cor}/c$, where $R_{\rm cor}\sim \mathcal{O}(10\,r_{\rm g})$ denotes the corona size and $r_{\rm g}$ is the gravitational radius, assuming cooling is inverse Compton dominated.

Consequently, the electron distribution remains mostly thermal at temperatures $k_BT_e \sim \mathcal{O}(100\,{\rm keV})$, albeit with possibly a small non-thermal fraction. The numerical simulation of the underlying microphysical processes has recently become accessible to particle-in-cell (PIC) codes incorporating radiative processes~\citep{groseljRadiativeParticleinCellSimulations2024,2024NatCo..15.7026N,groseljHighenergyEmissionTurbulent2026}. These simulations provide conclusive evidence that a strongly turbulent corona with characteristic Alfvén velocity $v_{\rm A}\gtrsim 0.1\,c$ and compactness in the expected range can reproduce the generic X-ray power law-like spectra, providing further support to the idea that these magnetised turbulent environments can also accelerate protons efficiently. However, contrary to the electron population, the proton distribution evolves on macroscopic timescales of the order of $R_{\rm cor}/v_{\rm adv}$, where $v_{\rm adv}\lesssim 0.1\,c$ denotes the flow velocity in the corona. Furthermore, simulating the acceleration of protons up to the energies inferred by IceCube would require a dynamical range covering at least five orders in magnitude. This problem is thus bound to remain out of reach of PIC simulations for a long time.  

The most promising, as well as standard, way to bridge this gap in scales is to track the evolution of particle distribution functions through macroscopic transport equations integrating models benchmarked on kinetic simulations as well as constraints on the ambient physical conditions derived from observations or general-relativistic magnetohydrodynamic (GRMHD) simulations. Steps in this direction have been taken in the aforementioned references, although often at the price of diverse approximations, for example, using semi-analytical accelerated proton spectra and/or radiative losses, a steady state description, and modelling acceleration with prescribed radiative losses. The problem indeed shares similar degrees of complexity with models of coronal X-ray emission, as it requires a time-dependent treatment of acceleration and of radiative losses, including the reprocessing of electromagnetic radiation. To address these issues, we introduce here a novel hybrid numerical framework, \texttt{Turb-AM3}, specifically designed to model stochastic particle acceleration in compact environments and to predict the ensuing multi-messenger signatures in a self-consistent manner. The code couples the time-dependent lepto-hadronic radiative solver \texttt{AM3} \citep{klingerAM$^3$OpenSourceTool2024} with a stochastic-acceleration module, which offers a detailed treatment of charged-particle energisation in turbulent plasmas. Particle acceleration is described through a momentum-space transport equation that incorporates recent theoretical developments, while the framework self-consistently evolves the magnetic turbulent cascade and includes a physically motivated description of the environment, including radiative (soft photon) field and plasma flow. The \texttt{AM3} module computes all relevant radiative and hadronic losses and tracks the production and propagation of secondary particles, including photons and neutrinos, carefully accounting for the development of electromagnetic cascade. This numerical framework is similar in spirit to that recently proposed in  \citet{Yuan2026}, but it contains additional features of interest that we discuss in this work. In particular, it includes a self-consistent description of the feedback exerted by accelerated particles on the magnetised turbulence and hence on the acceleration process, which is potentially significant in the present context given the inferred (high) proton luminosities. The code architecture and main principles are discussed in Sect.~\ref{sec:code}.

We used \texttt{Turb-AM3} to investigate the specific case of stochastic acceleration in the turbulent black hole corona of NGC~1068 and to derive self-consistent proton, photon, and neutrino spectra. Template neutrino spectra are of particular importance to both enhance the sensitivity of neutrino telescope searches and to constrain the overall contribution of black hole coronae to the diffuse neutrino background. As we show in the following sections of this work, our investigation also offers new insights on key open issues, such as the main physical parameters governing the neutrino spectral shape and luminosity, the effects of non-linear proton acceleration and radiation (e.g. impact on coronal turbulence and hadronic-induced pair production), and the expected characteristics of the lower-energy ($E \sim 1$--$100$ GeV) neutrino signal. Our model is self-consistent in that it treats the non-linear and time-dependent aspects of proton acceleration and neutrino production simultaneously, bridging the gap between PIC simulation scales and those of the source. In this mind set, we borrowed insights from recent PIC simulations to inform and improve our analysis. While the corona is modelled as a one-zone system in this study, the framework can be extended to state-of-the-art position-dependent physical parameters that can be derived from, for example, radiative GRMHD simulations. The Turb-AM3 framework is of broad applicability, and in the spirit of \texttt{AM3}~\citep{klingerAM$^3$OpenSourceTool2024} and similar codes~\citep[e.g.][and references therein]{2026ApJS..282...22C}, it could also be applied to other sources, such as hidden-neutrino sources, AGN jets~\citep{2024A&A...681A.119R}, X-ray binaries~\citep{2023MNRAS.524.1326K}, and/or tidal disruption events~\citep{2024ApJ...969..136Y} with suitable adjustments. 

This paper is organised as follows. We present the physical framework of the model in Sect.~\ref{sec:code}, starting with a generic description of the AGN corona, followed by the particle acceleration mechanism, the evolution of the turbulent cascade, and the numerical implementation of \texttt{Turb-AM3}. In Sect.~\ref{sec:results}, we apply our study to the case of NGC~1068 and discuss the multi-messenger signatures. We investigate the dependence on the (broad) parameter space in Sect.~\ref{sec:scan}, and extract template neutrino spectra for turbulent acceleration. In Sect.~\ref{otherseyferts}, we further show that the model provides a satisfactory match to the multi-messenger data of other Seyfert galaxies identified by IceCube (NGC~4151 and NGC~7469). We summarise our findings in Sect.~\ref{sec:conc}. Throughout this work, we use $Q_x/Q = 10^x$ in centimetre-gram-second (cgs) units unless we note otherwise.

\section{Physical and numerical set-up}\label{sec:code}
In this section we present the physical assumptions used to model particle acceleration, transport, and radiation in AGN coronae. Then we present the numerical implementation of the framework.

\subsection{Corona model}
\label{coronamodel}
We considered a general AGN configuration in which a SMBH of mass $M_{\rm BH}$, with gravitational radius $r_g \equiv GM_{\rm BH}/c^2$, is surrounded by an accretion disk of luminosity $L_{\rm d}$. In the case of NGC 1068, $M_{\rm BH}\simeq 10^{7.2}M_\odot$ and $L_{\rm d}\simeq5.0 \times 10^{44} \, \mathrm{erg \, s^{-1}}$ \citep{Woo_2002,lopez-rodriguezEmissionDistributionDust2018}. Above and below the disk, a hot, magnetised plasma forms a compact corona. Throughout this work the corona was assumed to be approximately spherical, with a characteristic size $R_{\rm cor} \lesssim 30\, r_g$, consistent with constraints for NGC~1068, especially the opacity needed for $\gamma\gamma$ absorption to be efficient \citep{dasRevealingProductionMechanism2024}. This approximation remains reasonable for the present purposes, as what truly matters is the effective size $R_{\rm cor}$, which controls the losses through diffusive escape, the overall radiative opacity and the energy density of radiative backgrounds.

The corona is moderately thick to Thomson scattering, with optical depth $\tau_\mathrm{T} = n_{e} R_{\rm cor}\, \sigma_\mathrm{T} \lesssim 1$ \citep{zdziarskiPowerlawXrayGammaray1985,sternGeometryXRayEmittingRegion1995,ricciBATAGNSpectroscopic2018},
in terms of $n_e$ the total electron and positron number density, including pairs produced in situ. The pair content of AGN coronae is not currently well constrained on observational grounds~\citep{2015MNRAS.451.4375F,ricciBATAGNSpectroscopic2018,2021MNRAS.506.4960H}. The pair loading factor, expressed as the ratio of proton to electron plus positron number densities ($n_p / n_e$), is regulated by the balance between photon-photon pair production and pair annihilation, and hence it depends on the photon and electron energy distributions~\citep[e.g.][]{1985ApJ...294L..79Z,1996ApJ...470..249P,2017MNRAS.467.2566F}. It can therefore be obtained only through detailed modelling of X-ray data, at the price of assumptions regarding the microphysics of electron energisation and of the photon backgrounds.  In this work, we adopted a fiducial pair-free configuration, $n_p / n_e = 1$, but discuss its influence on the parameters further below as well as in Sect.~\ref{sec:scan}; see also \cite{2025ApJ...995..166Y}.
Numerically, the electron number density is given by
\begin{align}
    n_e &\,\simeq\, 2\times 10^{10}\,{\rm cm}^{-3}\,\,\frac{\tau_{\rm T}}{0.5}\frac{R_{\rm cor}}{15\,r_{\rm g}} M_{7.2}\,.
    \label{eq:edens}
\end{align}

The corona is assumed to be predominantly powered by the dissipation of magnetic energy (e.g.~\citealt{1979ApJ...229..318G}). As the dissipated energy of the turbulence is converted into the X-ray spectrum on short timescales due to the high compactness 
~\citep{beloborodovRadiativeMagneticReconnection2017,groseljRadiativeParticleinCellSimulations2024,groseljHighenergyEmissionTurbulent2026}, the steady state energy density of X-ray radiation can be expressed as
\begin{align}
    u_{\rm X}&\,\approx\, \frac{t_\mathrm{\gamma,esc}}{t_{\rm diss}}\frac{\delta B^2}{8\pi},
    \label{eq:uX-diss}
\end{align}
where $t_\mathrm{\gamma,esc}\simeq (1+\tau_{\rm T})R_{\rm cor}/c$ represents the characteristic escape timescales of photons and $t_{\rm diss} \simeq f_{\rm diss}\ell_{\rm c}/v_{\rm A}$ is a typical time for dissipation of magnetic turbulence, with $\ell_{\rm c}$ as the coherence length of the turbulence.
The velocity $v_{\rm A} = \delta B/\sqrt{4\pi n_p m_p}$ is understood here as the Alfvénic velocity of magnetic fluctuations, which sets the characteristic velocity of the turbulent motions. We further assumed $f_{\rm diss}=0.1$, as dissipation takes several turn-around times to proceed; this value also corresponds to the reconnection rate, which provides the relevant scaling in relativistic turbulence~\citep{fiorilloMagnetizedStronglyTurbulent2024}. Noting that the X-ray luminosity $L_{\rm X} = V_{\rm cor} u_{\rm X}/t_\mathrm{\gamma,esc}$, where $V_{\rm cor}$ denotes the volume of the corona, and using Eqs.~(\ref{eq:edens}) together with (\ref{eq:uX-diss}), we obtained
\begin{align}
    \delta B&\,\simeq\, 4.8\times 10^3\,{\rm G}\,\,{L_{\rm X}}_{44}^{1/3}\left(\frac{R_{\rm cor}}{15\,r_{\rm g}}\right)^{-7/6} \left(\frac{\ell_{\rm c}}{5\,r_{\rm g}}\right)^{1/3}M_{7.2}^{-5/6}\,,\nonumber\\
    v_{\rm A} &\,\simeq\,0.2 c\,\,{L_{\rm X}}_{44}^{1/3}\left(\frac{R_{\rm cor}}{15\,r_{\rm g}}\right)^{-2/3} \left(\frac{\ell_{\rm c}}{5\,r_{\rm g}}\right)^{1/3}M_{7.2}^{-1/3}\,,
    \label{eq:B-vA}
\end{align}
with $M_{7.2}\equiv M_{\rm BH}/10^{7.2}M_\odot$, $L_{X_{44}}\equiv L_X/10^{44}\,$erg/s. The numerical estimates for $B$ and $v_{\rm A}$ assume $n_p/n_e=1$ and $f_{\rm diss}=0.1$; they scale as $f_{\rm diss}^{-1/3}(n_p/n_e)^{1/6}$ and $f_{\rm diss}^{-1/3}(n_p/n_e)^{-1/3}$ respectively. The corresponding plasma $\beta_p \equiv P_\mathrm{th}/(\delta B^2/8\pi)$ parameter is of the order of unity for a proton temperature $T_p \sim 2\times 10^{11}\,$K (close to the virial temperature at $R_{\rm cor}=15\,r_{\rm g}$). Following~\cite{2025PhRvL.135f5201G}, we note that $\beta_p\sim 1$ follows from the balance between turbulent dissipation and particle escape when $R_{\rm cor}/\ell_{\rm c}\sim O(1)$. Pair loading tends to increase the Alfvénic velocity and the proton temperature (assuming $\beta_p\sim 1$), yet it exerts a small influence on the overall turbulent magnetic energy density.

In the present paper, we do not model the physics of electron energisation, which is governed by fast processes acting on timescales well below those controlling proton acceleration, for example reconnection in microscopic current sheets~\citep[e.g.][]{beloborodovRadiativeMagneticReconnection2017}. Furthermore, a detailed modelling of the electron energy distribution requires detailed treatment of radiative interactions for mildly relativistic electrons~\citep[e.g.][]{1992MNRAS.258..657C,2008A&A...491..617B,2011MNRAS.414.3330V}, which we leave to a future study, as this does not otherwise impact the proton spectra that we seek to determine. We instead assumed that this distribution is characterised at all times by a Maxwellian with a dimensionless temperature $\Theta_e \equiv k_B T_e /(m_e c^2)\sim 0.2$.  Such a distribution in the rest frame of the central BH could also arise from Doppler broadening by strong turbulent motions, even if the electrons remain cool in the local co-moving frame of the turbulent plasma \citep{beloborodovRadiativeMagneticReconnection2017,2020ApJ...899...52S,groseljRadiativeParticleinCellSimulations2024}.

Our framework accounts for inverse Compton energisation of soft disk photons, that shapes the X-ray spectrum responsible for $p\gamma$ interactions at the origin of the high-energy neutrinos. To model the injection spectrum of soft photons in the corona, we approximate the accretion-disk emission with a multicolour blackbody spectrum, taking the spectral luminosity per unit energy as $L_\epsilon^{\rm disk} \propto \epsilon^{3}$ for $\epsilon < 0.1\,\epsilon_{\rm d}$ and $L_\epsilon^{\rm disk} \propto \epsilon^{4/3} \exp(-\epsilon/\epsilon_{\rm d})$ for $\epsilon \ge 0.1\,\epsilon_{\rm d}$, with a characteristic cut-off $\epsilon_{\rm d} = 31.5~\mathrm{eV}$. Due to geometrical and opacity effects, the corona intercepts only a fraction $f_{\rm cor} \sim 0.2$ of the disk luminosity $L_d$ \citep[e.g.][]{ghiselliniCanonicalHighpowerBlazars2009,dovciakPhysicalModelBroadband2022a}. 

This physical environment combines hot ions, sub-relativistic electrons, strong magnetic fields, dynamic turbulence, and a strong radiation field. Together, these conditions favour the stochastic acceleration of charged particles and the production of multi-wavelength and multi-messenger emission.

\subsection{Stochastic acceleration of protons}\label{sec:acc}
In magnetised turbulent coronae, stochastic proton acceleration mediated by repeated interactions between charged particles and the turbulent electric fields provides a natural pathway to acceleration. The efficiency of subsequent turbulent acceleration is governed by two factors, which remain insufficiently constrained today: the fraction of protons extracted from the thermal population and injected into the non-thermal component, and the properties of the turbulent energisation mechanism itself. Regarding the latter, stochastic particle acceleration is commonly described in the framework of a quasi-linear theory of resonant wave-particle interactions~\citep[e.g.][and references therein]{2002cra..book.....S} and numerically modelled using a purely diffusive Fokker-Planck scheme. However, this scheme appears to disagree with recent magnetohydrodynamic and PIC simulations (see e.g. \citealt{zhdankinKineticTurbulenceRelativistic2017,comissoInterplayMagneticallyDominated2019a,kimuraAccelerationEscapeProcesses2019,wongFirstprinciplesDemonstrationDiffusiveadvective2020a,trottaFastAccelerationTransrelativistic2020,bresciNonresonantParticleAcceleration2022,pezziRelativisticParticleTransport2022,2023ApJ...944..122M,puglieseEnergizationChargedTest2023,wongEnergyDiffusionAdvection2025}, and \citealt{lemoinePowerlawSpectraStochastic2020} for a discussion of these issues). This has spurred new theoretical ideas on modelling stochastic acceleration in strongly turbulent plasmas \citep[e.g.][]{lynnACCELERATIONRELATIVISTICELECTRONS2014,lemoineGeneralizedFermiAcceleration2019,lemoinePowerlawSpectraStochastic2020,demidemParticleAccelerationRelativistic2020,sioulasStochasticTurbulentAcceleration2020,sioulasSuperdiffusiveStochasticFermi2020,lemoineParticleAccelerationStrong2021,lemoineFirstPrinciplesFermiAcceleration2022,xuTurbulentReconnectionAcceleration2022,dasStudyingMirrorAcceleration2025a,2025PhRvE.112a5205L}. Here we remain agnostic and envisage several prescriptions for modelling transport in momentum space, including the standard Fokker-Planck approach employing a diffusion coefficient extracted from PIC simulations and a generalised description of Fermi acceleration~\citep{lemoineFirstPrinciplesFermiAcceleration2022}.

In relativistic turbulence, magnetic reconnection in microscopic current sheets acts as the injection mechanism~\citep{comissoInterplayMagneticallyDominated2019a}. It extracts a significant fraction of protons from the thermal pool and pre-accelerates them before they undergo stochastic energisation by turbulent magnetic fields. The turbulent magnetisation parameter is defined as $\sigma_{\delta B} = \delta B^2/(4\pi n_pm_pc^2)$, also written as the Alfvén 4-velocity squared $\sigma_{\delta B} = (v_{\rm A}/c)^2/[1-(v_{\rm A}/c)^2]$. For reference, \citet{comissoInterplayMagneticallyDominated2019a} report an injected particle fraction of $\xi_{n_p} \sim 0.2-0.3$ at $\sigma_{\delta B}\sim 10$, where $\xi_{n_p} \equiv n_{p,\mathrm{nth}}/n_{p,\mathrm{tot}}$ denotes the number fraction of particles injected into the non-thermal population. In terms of pressure ratio between non-thermal and thermal particles, hereafter $\xi_p \equiv (P_{p,\mathrm{nth}}/P_{p,\mathrm{tot}})$, this implies $\xi_p$ close to unity. However, these  fractions, and possibly the injection mechanism itself, may vary in mildly or sub-relativistic regimes~\citep{comissoIonElectronAcceleration2022a,wongEnergyDiffusionAdvection2025}. Consequently, 
within our mildly relativistic framework ($\sigma_{\delta B} \sim 0.1$), we treated the injection  
fraction $\xi_p$ as a free parameter. We adopted a fiducial value of  $\xi_p = 0.1$ and assumed that particles are injected at $\epsilon_{\mathrm{p,inj}} = \mathcal{O}(m_p c^2)$. The key parameter is here $\xi_p$, not $\epsilon_{\mathrm{p,inj}}$, whose exact value does not exert a strong influence on the final spectra. Since $\xi_p/\xi_{n_p} \simeq \epsilon_{\mathrm{p,inj}}/k_B T_p$, this choice of $\xi_p$ corresponds to $\xi_{n_p} \sim 10^{-2}$. We discuss the influence of $\xi_p$ on our results in Sect.~\ref{sec:scan}. 

The acceleration of the non-thermal proton spectral density $\mathcal{N}_{p} \equiv dN_p/dp = 4\pi p^2 f(p,t)$ (with $f$ the distribution function and $N_p$ the number density of protons) is modelled with a transport equation including an operator describing the stochastic acceleration, whose form depends on the physical regime considered. Here, we focus on the differential operator modelling this evolution in momentum space, and discuss in Sect.~\ref{sec:transport} the full transport equation for protons (and other species).

For small-amplitude quasi-linear type interactions, we followed standard treatments and described the diffusion process using a purely diffusive Fokker-Planck equation for momentum $p$,
\begin{equation}
    \mathcal{L}_{\rm FP} \mathcal{N}_p 
    \equiv \partial_p\!\left(D_{pp}\,\partial_p \mathcal{N}_p\right)
    - 2\,\partial_p\!\left(\frac{D_{pp}}{p}\, \mathcal{N}_p\right),
    \label{diffusive}
\end{equation}
which is characterised by the diffusion coefficient $D_{pp}$. We also used $D_{pp} = 0.3\,\sigma_{\delta B}\,p^2\, c/\ell_{\rm c}$ as obtained from numerical PIC simulations in the relativistic regime $\sigma_{\delta B}\gtrsim 1$. Recently, \citet{wongEnergyDiffusionAdvection2025} reported a different scaling in the sub-relativistic regime $\sigma_{\delta B}<1$, $D_{pp}\propto \sigma_{\delta B}^{3/2}$ and confirmed  that in order to reproduce the observed energy distributions, one needs to introduce a net advection coefficient with a non-trivial energy dependence. In light of this discrepancy, and noting that an order of unity modification in the diffusion coefficient translates in a similar renormalisation of the magnetisation, we retain the above value of $D_{pp}$ and compare in Appendix~\ref{sec:app} our different schemes of stochastic acceleration to evaluate the impact on the neutrino spectra. In this regime, the acceleration rate is given by $\nu_{\mathrm{acc}}(p) = t_{\rm acc}(p)^{-1} = 4D_{pp}/p^2$.

In strongly turbulent coronae, where $\delta B/B_0\gtrsim1$ ($B_0$ background field, $\delta B$ characteristic turbulent amplitude), the acceleration process is influenced by the intermittency of the turbulent fluctuations.
To describe this regime, we relied on the transport equation introduced in \citet{lemoineFirstPrinciplesFermiAcceleration2022}, which was successfully benchmarked on the results of magnetohydrodynamic simulations at $v_{\rm A}/c\simeq 0.4$, close to the range that we investigate. The corresponding differential operator reads as
\begin{equation}
    \mathcal{L}_{\rm GF} \mathcal{N}_p \equiv \int_0^\infty 
    2\pi c\left[
        \mathcal{N}_{p'}\, \frac{\varphi(p|p')}{l_g(p')}
        - \mathcal{N}_{p}\, \frac{\varphi(p'|p)}{l_g(p)}
    \right] dp',
    \label{master_like}
\end{equation}
where $\varphi(p|p')$ is the transition probability of momentum $p'$ to jump to momentum $p$ and $l_g(p) \equiv 2\pi pc/eB$ is the gyroradius scale of the particle. This master-type operator allows for non-local transfers in momentum space and captures regimes where interactions induce finite or large momentum jumps. It describes a `generalised Fermi' process in which protons can gain or lose energy as they cross intermittent regions of dynamic, curved and/or compressed magnetic field lines. In this formalism, energisation results from the crossing of velocity gradients, $\Gamma_{l_g}$, proceeding according to $\dot p = \Gamma_{l_g} p$, implying an acceleration rate of $\nu_{\mathrm{acc}}(p) = t_{\mathrm{acc}}(p)^{-1} = \langle\Gamma_{l_g(p)}\rangle$. For a recent application to the modelling of the spectral energy distribution (SED) of microquasars, see~\cite{2026arXiv260309394D}.

If a substantial fraction of the turbulent energy is dissipated into the proton non-thermal component, as is suggested by the high neutrino luminosity inferred from IceCube observations of Seyfert galaxies, the energy lost by the cascade must be accounted for. This backreaction process alters the cascade, the acceleration process, and thus eventually the proton spectra. This implementation is discussed in Sect.~\ref{sec:fdbck}.

\subsection{Transport equation for various species}\label{sec:transport}

\begin{table*}[t]
\centering
\caption{Physical processes included in the transport equation for each particle species for our framework.}
\label{tab:transport_terms}
\begin{tabular}{lcccc}
\hline\hline
Particle 
& Stochastic acceleration $\mathcal{L}_{\mathrm{stoch}}$
& Energy losses $\partial_p(p\,\nu_{\mathrm{loss}} N)$
& Escape $N/t_{\mathrm{esc}}$
& Source term $Q$
\\
\hline
Protons 
& Yes ($\mathcal{L}_{\mathrm{FP,GF}}$)
& $pp$, $p\gamma$, Bethe-Heitler (BH)
& Diffusion
& Injection
\\ \hdashline
Leptons ($e^-$,$e^+$) 
& Neglected
& Synch., IC
& Diffusion
& $e^-$ : Injection,
\\
& 
&
& 
& $e^\pm$ : $\gamma\gamma$/BH pair creation
\\ \hdashline
Photons 
& -
& -
& Eff. light crossing,
& Disk radiation,
\\

&
&
& $\gamma\gamma$ absorption
& radiative emission
\\ \hdashline
Neutrinos 
& -
& -
& Light crossing
& $\pi^\pm$, $\mu^\pm$ decays
\\ \hdashline
Pions and Muons 
& Neglected
& Neglected
& Disintegration
& Hadronic interactions
\\
\hline
\end{tabular}
\end{table*}

For each particle species ($s$) considered here (mainly protons, leptons, photons, and neutrinos but also pions and muons), the evolution of the particle density spectrum $\mathcal{N}_s \equiv dN_s/dp$ is described by a transport equation of the form
\begin{equation}
    \partial_t \mathcal{N}_s
    = \left[\mathcal{L}_{\mathrm{stoch}} \mathcal{N}_s\right]_{s=p} 
    - \partial_p\!\left(p\,\nu_{\mathrm{loss}}\, \mathcal{N}_s\right)
    - \frac{\mathcal{N}_s}{t_{\mathrm{esc},s}}
    + Q_s\,
    \label{eq:transport}
\end{equation}
with the following definitions. First, the terms in brackets only apply for protons, as they represent stochastic acceleration; as explained in Sect.~\ref{sec:loss}, we do not treat the dynamical evolution of the electron and positron populations, which evolve on short timescales. The operator $\mathcal{L}_{\mathrm{stoch}}$ corresponds to the differential operator $\mathcal{L}_{\rm FP}$ or $\mathcal{L}_{\rm GF}$ described earlier. The second term on the right hand side describes systematic momentum losses due to interactions through the rate $\nu_{\mathrm{loss}}$, further discussed in Sect.~\ref{sec:loss}. The third term parametrises particle removal through diffusive escape from the corona (Sect.~\ref{sec:esc}), annihilation (for pairs and photons), hadronic interactions (for protons), or decay (pions and muons). The source term $Q_s$ accounts for particle injection (for protons and electrons) or production through interactions, for example annihilation (pairs and photons) and radiative processes (pions, muons  and neutrinos). These various processes and their corresponding implementation are described in detail in \citet{klingerAM$^3$OpenSourceTool2024}. Depending on the particle species and the physical processes involved, some of these terms may be absent or take different explicit forms. A summary of the transport terms retained for each species in our model is given in Table~\ref{tab:transport_terms}. 

As in \citet{lemoineNeutrinosStochasticAcceleration2025}, we integrate the above transport equations up to time $t_{\rm adv}$, which characterises the crossing time of the corona by advection with the background plasma at velocity $v_{\rm adv}$, as described in Sect.~\ref{sec:esc}.
This describes an inflow of the corona plasma but it could also describe an outflow, for example in a jetted corona.
The overall proton spectral shape is governed by the influence of radiative losses (mostly at the highest energies), the consequence of turbulent damping, and the relative importance of acceleration, escape, and advection processes. This interplay can be conveniently characterised by the ratios of timescales
\begin{equation}
    \frac{t_{\mathrm{esc}}}{t_{\mathrm{acc}}}
\approx \left(\frac{R_\mathrm{cor}}{\ell_c}\right)^2 \frac{v_\mathrm{A}}{c}\,,
\qquad
    \frac{t_{\mathrm{adv}}}{t_{\mathrm{acc}}}
\approx \frac{R_\mathrm{cor}}{\ell_c}\,
\frac{v_\mathrm{A}^2}{v_\mathrm{adv} c}
\,.
    \label{eq:def-Xi}
\end{equation}
Large values of $t_{\mathrm{esc}}/t_{\mathrm{acc}}$ imply efficient confinement, allowing protons to undergo multiple acceleration cycles and build up a hard, extended non-thermal spectrum. Conversely, when $t_{\mathrm{esc}}/t_{\mathrm{acc}} \ll 1$, escape dominates over acceleration, leading to a suppressed high-energy proton population and neutrino production. The second ratio, $t_{\mathrm{adv}}/t_{\mathrm{acc}}$, controls the maximum energy that protons can achieve before being advected out of the corona, into the black hole or in the jet. For $t_{\mathrm{adv}}/t_{\mathrm{acc}} \ll 1$, advection removes particles faster than they can be accelerated, and the proton spectrum exhibits a high-energy cut-off controlled by advection, i.e. by the maximum time over which acceleration can take place. In contrast, when $t_{\mathrm{adv}}/t_{\mathrm{acc}} \gtrsim 1$, protons have sufficient time to reach energies at which radiative and photo-hadronic losses become dominant.

\subsection{Radiative and hadronic losses and timescales}\label{sec:loss}
\begin{figure}
\resizebox{\hsize}{!}{\includegraphics[clip=true]{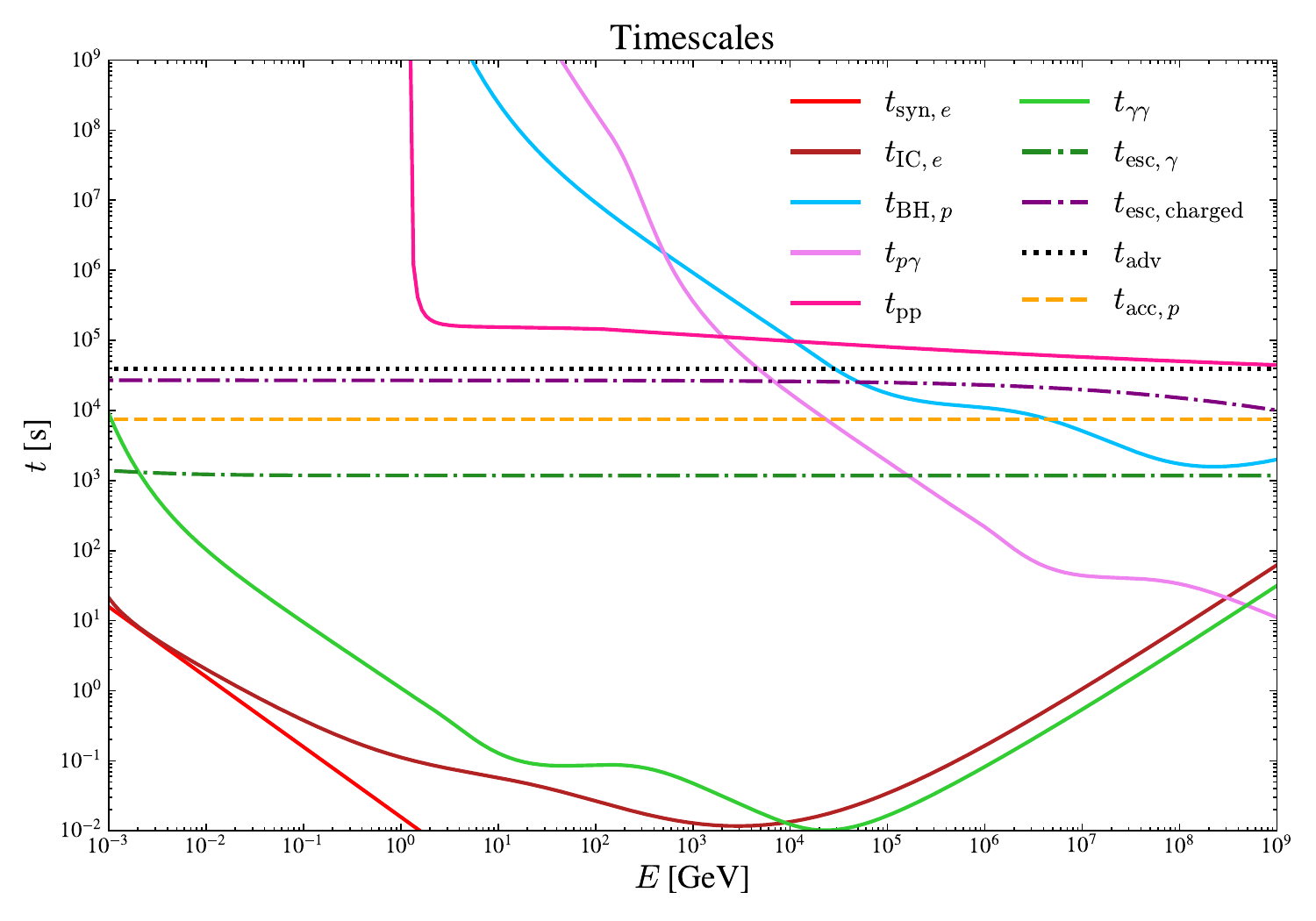}}
\caption{Timescales of the different processes occurring in the AGN corona, including the escape of charged and neutral particles, the acceleration of protons, radiative and hadronic losses of protons, electrons and positrons, $\gamma\gamma$ annihilation of photons, and the advection timescale of the coronal plasma.}
\label{fig:timescales}
\end{figure}

Particles experience systematic energy losses through radiative and hadronic processes. The relevant timescales are shown in Fig.~\ref{fig:timescales} for protons, electrons, and photons, as computed with \texttt{AM3} for a fiducial configuration in which the disk luminosity is $L_d = 5.0 \times 10^{44}\,\mathrm{erg\,s^{-1}}$, the corona has a characteristic size $R_{\rm cor} = 15\,r_g$, $\tau_\mathrm{T} = 0.5$, $n_p/n_e = 1$, $\xi_p = 0.1$, $\ell_c = 4.0 \,r_g$, and $v_\mathrm{A} = 0.25 \, c$. For comparison, the escape and acceleration timescales are also displayed.

Protons and electron-positron pairs in the corona lose energy through interactions with the ambient magnetic and radiation fields, as well as through hadronic and other plasma processes. 
Given the compactness $\ell_{\rm comp} \equiv 4\pi (m_p/m_e) (L_{\rm d}/L_{\rm edd})(r_{\rm g}/R_{\rm cor})\simeq 50\,L_{\rm d,44} (R_{\rm cor}/15\,r_{\rm g})^{-1}M_{7.2}^{-1}$, with $L_{\rm edd}$ the Eddington luminosity, and given the plasma magnetisation $\sigma_{B}\simeq 0.1 (B/5\times 10^3\,{\rm G})^2(R_{\rm cor}/15\,r_{\rm g})x_p^{-1}(\tau_{\rm T}/0.5)^{-1}$, we see that the pairs cool efficiently through synchrotron and inverse Compton losses. In detail, inverse Compton losses of electrons with Lorentz factor $\gamma_e=\varepsilon_e/m_ec^2$ occur on $t_{\rm IC,e} = 3R_\mathrm{cor}/(4\gamma_e\ell_\mathrm{comp} c) \simeq 0.02 (R_{\rm cor}/c)\, (\ell_\mathrm{comp}/50)^{-1} (\varepsilon_e/600~{\rm keV})^{-1}  \ll R_\mathrm{cor}/c$, and similarly for synchrotron losses, $t_{\rm syn, e} = (3m_e/2m_p)R_\mathrm{cor}/(\gamma_e\sigma_{B} c) \simeq 0.02  (R_{\rm cor}/c)\,(\sigma_B/0.1)^{-1}(\varepsilon_e/600~{\rm keV})^{-1} \ll R_\mathrm{cor}/c$.  This justifies our approximation of fixing the electron distribution function to its steady state Maxwellian shape on the timescales of proton evolution, that are larger than $R_{\rm cor}/c$. For comparison, the typical acceleration timescale for stochastic acceleration reads as
\begin{align}
    t_{\rm acc} \simeq 4\, \frac{R_{\rm cor}}{c}\,\left(\frac{v_\mathrm{A}}{0.25c}\right)^{-2} \left(\frac{\ell_c}{5r_g}\right) \,.
    \label{eq:tacc}
\end{align}

Protons within the corona lose momentum through the following radiative and hadronic 
processes: synchrotron radiation, inverse Compton scattering, inelastic 
proton-proton ($pp$) collisions, photo-hadronic ($p\gamma$) interactions, and 
Bethe-Heitler pair production (BH). However, proton synchrotron and inverse Compton losses can be neglected in the conditions of AGN coronae, as their timescales are $(m_p/m_e)^4$ larger than the corresponding electron loss  timescales. Proton  losses are dominated by the hadronic ($pp$, $p\gamma$) and Bethe-Heitler channels~\citep{muraseHiddenCoresActive2020,eichmannSolvingMultimessengerPuzzle2022,dasRevealingProductionMechanism2024}. More specifically, our detailed numerical computations using \texttt{AM3} indicate that  $pp$ interactions dominate for protons below 1~TeV and that BH pair production and $pp$ and $p\gamma$ 
interactions are all relevant in the range between 1 and 10~TeV, while $p\gamma$ eventually dominates above 10~TeV (see Fig.~\ref{fig:timescales}). For our fiducial parameters, the proton cooling timescale falls below the acceleration timescale at energies $\gtrsim 10~\mathrm{TeV}$, as required to produce the $1$–$10~\mathrm{TeV}$ neutrinos observed
by IceCube. For photons, the $\gamma\gamma$-annihilation timescale computed with \texttt{AM3}, 
shown in Fig.~\ref{fig:timescales}, becomes shorter than the photon escape 
timescale for energies above $\sim 100~\mathrm{keV}$. High-energy gamma-ray photons are thus efficiently absorbed and reprocessed within the corona. In contrast, neutrinos escape  freely, so their flux remains unaffected and can exceed the 
observed photon flux at TeV energies. Finally, pion and muon energy losses and acceleration can be safely neglected,
as their typical decay timescales are much shorter than any radiative or 
hadronic cooling timescale.

\subsection{Spatial transport in the corona}\label{sec:esc}
Protons, electrons, positrons, and other charged particles can escape from the corona on a characteristic diffusive timescale $t_{\pm,\rm esc} = R_{\rm cor}^2/(2\kappa)$, with spatial diffusion coefficient $\kappa$. This coefficient includes both turbulent diffusion by advection with the large-scale eddies and pitch-angle scattering on magnetic inhomogeneities in the co-moving frame. The turbulent contribution is approximated as $\kappa_{\rm turb} \simeq \ell_c v_\mathrm{A}/3$, in agreement with recent PIC simulation results \citep[][]{groseljHighenergyEmissionTurbulent2026}, while the scattering contribution is taken as $\kappa_{\rm scatt} \simeq r_L^{1/3}\ell_c^{2/3}c/3$ (see e.g. \citealt{berezinskiiAstrophysicsCosmicRays1990} and  \citealt{2002cra..book.....S}; for recent studies of large amplitude turbulence, see \citealt{kempskiCosmicRayTransport2023} and  \citealt{lemoineParticleTransportLocalized2023}). The total diffusion coefficient, $\kappa = \kappa_{\rm turb} + \kappa_{\rm scatt}$, therefore leads to a momentum-dependent escape timescale for charged particles.

We also assumed that the plasma is advected through the corona with the characteristic velocity $v_{\rm adv}$. This transport characterises advection into the central black hole, but it can also model advection along an outflow in models where the corona lies on a jet boundary~\citep{2025ApJ...979..199S}. We assumed a velocity describing radial advection \citep[see][]{shakuraBlackHolesBinary1973, frankAccretionPowerAstrophysics2002}
\begin{equation}
    v_\mathrm{adv}(r) \simeq \,\alpha\, v_\mathrm{K}(r)
    = \,\alpha\, c\, \hat{r}^{-1/2},
\end{equation}
where $\alpha \simeq 0.1$ is the accretion viscosity parameter, $v_\mathrm{K}(r)$ is the Keplerian velocity, and $\hat{r} \equiv r/r_g$.  For a corona of size $R_\mathrm{cor} = 15\,r_g$, the advection speed at the outer boundary typically reaches $v_\mathrm{adv}(R_\mathrm{cor}) \simeq 0.03\,c$.

Protons were assumed to be continuously injected at the outer edge of the corona and were subsequently transported by advection until they are advected out of the corona. In this configuration, the temporal evolution of a particle population is directly mapped onto its radial evolution through $dt = dr/v_\mathrm{adv}(r)$ so that the time-dependent spectrum of a particle can equivalently be described as a function of its radial position in the corona. As discussed in \citet{lemoineNeutrinosStochasticAcceleration2025}, this method also provides a satisfactory model of the case in which protons are continuously injected at all points in the corona, at least as long as the corona is treated as one-zone.

In this work, we adopted the simplifying approximation of a constant advection velocity throughout the corona, $v_\mathrm{adv}(r) \simeq v_\mathrm{adv}(R_\mathrm{cor})$. Under this assumption, the advection timescale reduces to $t_\mathrm{adv} \equiv R_\mathrm{cor}/v_\mathrm{adv}$ so that the spatially averaged proton spectrum can be expressed as a simple time average,
\begin{equation}
\mathcal{N}_\mathrm{p,cor} = 
    \frac{1}{t_\mathrm{adv}}
    \int_{0}^{t_\mathrm{adv}} \mathcal{N}_p(t)\, dt\,.
    \label{eq:mean_population_const_v}
\end{equation}
The distributions $\mathcal{N}_p(t)$ are obtained as the time-dependent solutions of the transport equation Eq.~(\ref{eq:transport}).\footnote{Equation~(\ref{eq:mean_population_const_v}) implicitly assumes that particles are confined in, and advected with the plasma in ballistic motion, i.e $t_{\rm esc}\gg t_{\rm adv}$. It remains valid in the opposite limit $t_{\rm adv}\gg t_{\rm esc}$, when advection does not play any significant role. In such a case, the limit $t_{\rm adv}\rightarrow +\infty$ can be taken, and Eq.~(\ref{eq:mean_population_const_v}) then provides the steady-state solution to the transport equation. In the intermediate limit $t_{\rm adv}\sim t_{\rm esc}$, it remains a satisfactory approximation, as we have checked using a simple Monte-Carlo model describing spatial transport, neglecting particle energisation.}

Although the assumption of a constant advection velocity is clearly idealised, its impact on the results is partially mitigated by the fact that a stronger turbulence and magnetic amplification closer to the black hole are expected to increase the Alfvén speed, $v_\mathrm{A}$, which would in turn boost the acceleration rate. This effect acts in the opposite direction to the increase of $v_\mathrm{adv}$ close to the black hole, and the two effects tend to partially compensate each other. Detailed information from radiative GRMHD simulations of accreting black holes would be highly valuable in that respect, and could be incorporated into the present framework. The advection timescale  (Fig.~\ref{fig:timescales}) sets an upper limit on the available time for acceleration, unlike $t_{\rm esc}$, which characterises a stochastic process of escape.

\subsection{Turbulent cascade and damping by energetic particles}\label{sec:fdbck}
Matching the neutrino flux observed by IceCube requires a non-thermal proton energy density comparable to the magnetic energy density in the corona~\citep{dasRevealingProductionMechanism2024}. Furthermore, the energy fraction of the non-thermal population $\xi_p$ inferred in relativistic \citep{comissoInterplayMagneticallyDominated2019a} and mildly relativistic \citep{comissoIonElectronAcceleration2022a} PIC simulations indicate that the energy density in non-thermal protons can become comparable to the magnetic pressure. Under these conditions, the dissipation of turbulent energy into particle acceleration becomes potentially significant with regards to the flow of turbulent energy through the cascade, implying substantial damping of turbulent power. A self-consistent description of the dissipation of turbulent energy and its impact on particle acceleration is therefore an important and well-motivated ingredient of our model.

The co-evolution of stochastic acceleration and turbulence dissipation has been addressed in various astrophysical contexts in recent years~\citep[e.g.][]{2016ApJ...816...24K,2022MNRAS.517.2502S,2025ApJ...989...99G}. To describe this non-linear interplay, we followed the general treatment of \citet{lemoineNonlinearAspectsStochastic2024} and \citet{lemoineNeutrinosStochasticAcceleration2025}. We assumed that the turbulent fluctuations are  characterised by a power spectrum $S_k \propto k^{-5/3}$ in terms of wavenumber $k$, with $\mathcal{E}_{B_k} \propto k S_k \propto k^{-2/3}$, and overall normalisation  $\int {\rm d}\ln k \,\,\,\mathcal{E}_{B_k}\,=\,\delta B^2/8\pi$. Energy is injected at the correlation scale $\ell_c \sim k_{\rm inj}^{-1}$ and cascades through the inertial range ($k_{\rm inj} \ll k \ll k_{\rm diss}$) with a constant flux $\gamma_k \mathcal{E}_{B_k}$, where $\gamma_k \propto k^{2/3}$.
To model the dissipation of turbulent power -- strictly speaking, of the turbulent electric fields -- by particle acceleration, we introduce a kernel $\phi(k,p)$ that specifies how turbulence at wavenumber $k$ feeds particles of momentum $p$. The cascade equation is therefore
\begin{equation}
    \partial_t \mathcal{E}_{B_k}
    = - k \partial_k (\gamma_k \mathcal{E}_{B_k})
      - \int \phi(k,p) \, \varepsilon_p p\,\partial_t \mathcal{N}_p(p,t)\, d\ln p,
\label{eq:cascade_damped}
\end{equation}
guaranteeing overall energy conservation between the proton non-thermal energy distribution and the turbulence spectrum at a given point, thanks to the normalisation property $\int{\rm d}\ln k\,\phi(k,p)\,=\,1$. For simplicity, we assumed that $\phi(k,\,p)$ describes 
interactions at $k\sim (eB/pc)$, as characterised by a Gaussian centred at $(\ln k + \ln(pc/eB))
$ and normalised to unity. This choice does not affect our conclusions, as discussed in the above references. Conversely, the damping of the cascade reduces the efficiency of stochastic acceleration, as modelled through the time-dependent acceleration rate:
\begin{equation}
   \nu_{\rm acc}(p,\,t) =  \nu_{\rm acc}(p,\,t=0)\, a 
           \int \phi(k,p)\, \mathcal{E}_{B_k}(t)\, d\ln k,
    \label{eq:eff.accel}
\end{equation}
with $a =\left(\int \phi(k,p)\, \mathcal{E}_{B_k}(t=0)\, d\ln k \right)^{-1}$ as a normalisation coefficient. The time-dependent acceleration rate enters Eq.~(\ref{eq:transport}) through the diffusion coefficient or the velocity gradients, as discussed in Sect.~\ref{sec:acc}.

In practice, damping becomes significant when the power injected into particles competes with the turbulent energy flux. Utilising the acceleration frequency $\nu_{\rm acc} \simeq v_\mathrm{A}^2/(c \ell_c)$ and the cascade rate $\gamma_{k_{\rm inj}} \simeq f_{\rm diss}\,v_\mathrm{A}/\ell_c$, and substituting $\mathcal{E}_{B_{k_{\rm inj}}} \simeq \delta B^2/(8\pi) = P_\mathrm{th}/\beta_p$, the condition for the onset of damping reads as
\begin{equation}
    u_{p,\rm non-th}(t) \sim \frac{f_{\rm diss}}{v_{\rm A}/c}\frac{P_\mathrm{th}}{\beta_p}.
    \label{eq:Pnth-selfr}
\end{equation}
Once damping sets in, proton acceleration slows down. In practice, if the cascade is quenched at wavenumber $k_{\rm damp}$, only particles with gyroradius $r_{\rm L} = pc/eB > k_{\rm damp}^{-1}$ continue to be accelerated, and their energy gain is limited by the turbulent energy available at $k_{\rm damp}$. Beyond that point,  the proton energy spectrum flattens out to equal energy per decade, meaning approximately ${\rm d}N_p/{\rm d}\varepsilon_p\propto \varepsilon_p^{-2}$. As a result, the maximum non-thermal energy density is here limited to a characteristic value close to the turbulent and plasma pressure.

\subsection{Numerical set-up}
The numerical strategy implemented in \texttt{Turb-AM3} aims at solving  self-consistently the momentum-space transport equation for protons  (Eq.~\ref{eq:transport}) together with the turbulent cascade equation  (Eq.~\ref{eq:cascade_damped}), while all other particle species (electrons,  positrons, photons, neutrinos, muons, and pions) are evolved with the standard \texttt{AM3} solver, Eq.~(\ref{eq:transport}) without the acceleration term. 

The physical state of the corona is specified by a set of characteristic  parameters: the mass of the central SMBH $M_\mathrm{BH}$,  the coronal radius $R_{\rm cor}$, the bolometric luminosity of the accretion disk $L_d$, the magnetic-field strength $\delta B\sim B$, the  turbulence coherence length $\ell_c$, and the advection velocity $v_\mathrm{adv}$.  The electron density $n_e$ is chosen such that the optical depth is of order 
unity, then the proton density is determined via the proton-to-lepton  density ratio $n_p/n_e$. We also specify the proton and electron  temperatures, $T_p$ and $T_e$, and the initial energy fraction between non-thermal and total protons,  $\xi_p$. The stochastic acceleration mechanism implemented in the simulation is  determined by selecting the appropriate transport operator from among  $\mathcal{L}_{\rm FP}$ and $\mathcal{L}_{\rm GF}$; the distribution function can also be evolved by a pure advection process characterising Fermi-I acceleration, but we do not consider it here. Finally, we define numerical quantities including the grid in momentum space and the timestep, the latter of which is taken as a fraction of the characteristic light-crossing time $\ell_c/c$.

The initial particle populations are then constructed as follows. Electrons are injected following a thermal Maxwell-Jüttner distribution at a temperature $T_e$, normalised to the electron density $n_e$. Non-thermal protons are initialised with a power-law shape $dN_p/dp \propto 
p^{-4}$ as observed for stochastic acceleration in the sub-relativistic regime~\citep{comissoIonElectronAcceleration2022a,lemoineFirstPrinciplesFermiAcceleration2022}, with $p_\mathrm{min} = m_pc$ and $p_\mathrm{max} = 5\times 10^8 \, \rm GeV$, but the precise parametrisation does not have an impact on the final proton spectrum. Their initial total pressure is normalised to $\xi_p P_\mathrm{th}$. The photon field  is initialised as a multi-temperature blackbody spectrum representing the  disk radiation field. While the injection remains constant in time, the 
photon population evolves dynamically through all included radiative,  hadronic, and pair-production processes.

At each timestep, the non-thermal proton distribution is advanced by computing a full step of stochastic acceleration (Eq.~\ref{eq:transport} without escape, injection, and losses), and turbulent damping (Eq.~\ref{eq:cascade_damped}). This requires solving the modified turbulent cascade equation, including backreaction from the energetic particles through the kernel $\phi(k,p)$, and computing the associated effective acceleration rate defined by Eq.~(\ref{eq:eff.accel}). The resulting proton spectral energy density is interpolated from momentum space onto the \texttt{AM3} energy grid. The \texttt{AM3} radiative solver then advances all species, using the updated proton spectrum as input, and taking into account losses and escape, Eq.~(\ref{eq:transport}) without the acceleration term. The updated spectra of all species are stored, and the proton spectrum is passed to the next acceleration step.

The electron population is taken to be constant in time, as it evolves on such short timescales that a steady state is a satisfactory approximation. For photons, we emphasise that initially only the disk photon distribution is an input. The X-ray component is a consequence of the inverse Compton scattering of the soft photons of the disk by the sub-relativistic thermal electrons, and is computed with the AM3 code. 

The duration of one simulation is $t_\mathrm{adv}$. The average particle populations of all species over the corona volume are computed as in Eq.~(\ref{eq:mean_population_const_v}). This average represents the stationary spectrum of the corona, and must be distinguished from the time-dependent spectra of each particle species at different points in the corona.

\section{Multi-messenger emission from NGC1068}\label{sec:results}
\subsection{Fiducial parameter set}\label{sec:fiducial}

\begin{figure*}
\resizebox{\hsize}{!}{\includegraphics[clip=true]{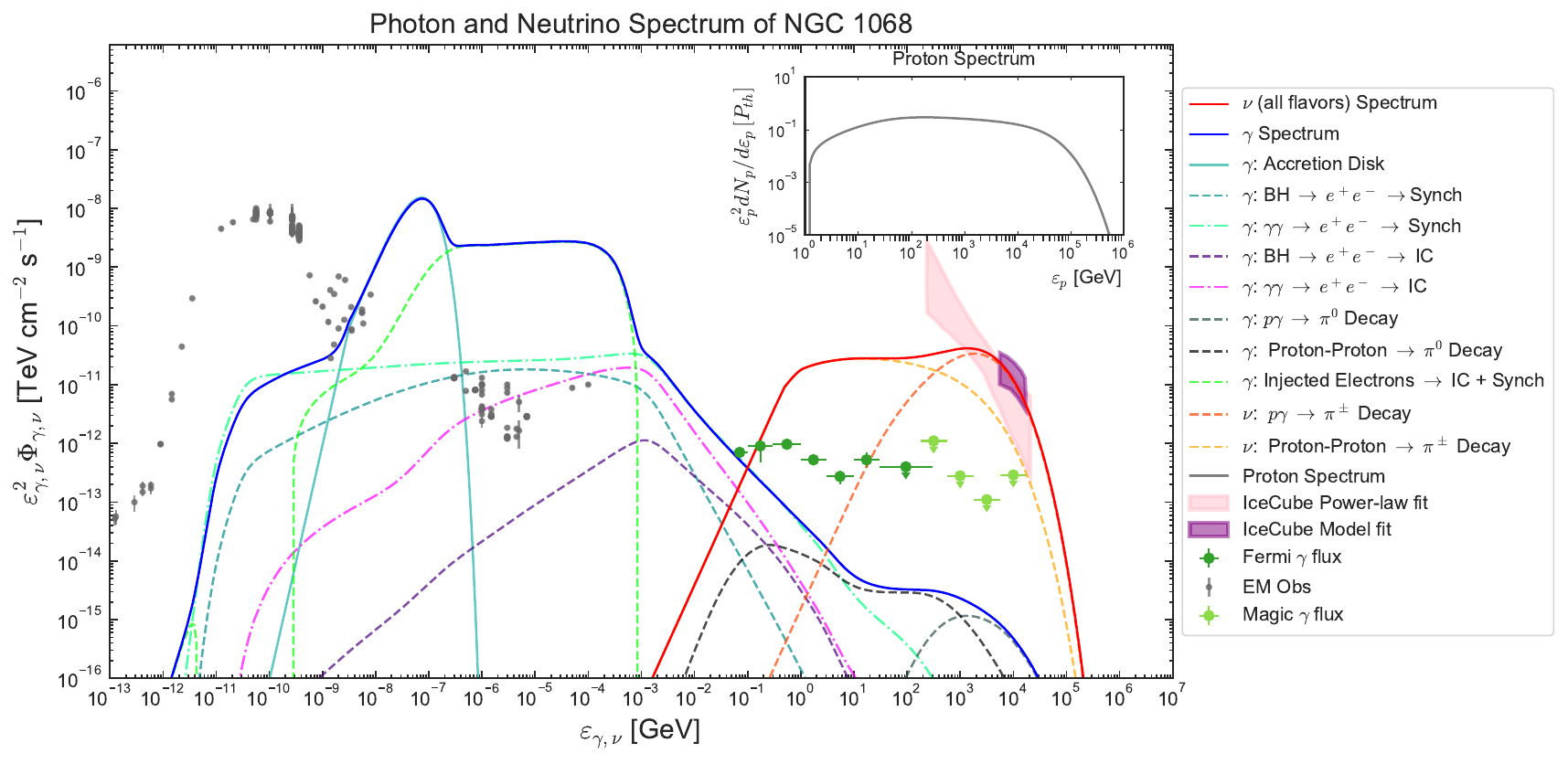}}
\caption{
Photon (blue) and neutrino (red) spectra from NGC~1068 computed with \texttt{Turb-AM3} for a fiducial set of parameters: $L_d = 5.0 \times 10^{44} \, \mathrm{erg \, s^{-1}}$, $R_\mathrm{cor} = 15 \, r_g$, $\xi_p = 0.1$, $\ell_c = 4.0 \, r_g$, $v_\mathrm{adv} = 0.03 \, c$, and $v_\mathrm{A} = 0.25 \, c$. The contributions from the different radiative and interaction processes are shown separately. BH denotes Bethe–Heitler pair production, $\gamma\gamma$ indicates photon–photon self-absorption, and IC refers to inverse Compton scattering. 
All spectra are shown as fluxes observed at Earth, assuming a luminosity distance of $10\,\mathrm{Mpc}$. Neutrino data from IceCube \citep{abbasiEvidenceNeutrinoEmission2025} and electromagnetic observations \citep{changOpenUniverseVOUBlazars2019,thefermicollaborationFermiLargeArea2020,magiccollaborationConstraintsGammarayNeutrino2019} are included for comparison.
The stationary proton spectrum obtained with \texttt{Turb-AM3} is displayed in grey in the inset, where the energy distribution is normalised to the background plasma pressure $P_\mathrm{th}$.
}
\label{fig:proton_photon_neutrino_stat}
\end{figure*}
We applied \texttt{Turb-AM3} to compute the photon and neutrino spectra from NGC~1068 for our 
fiducial set of parameters: $L_d = 5.0 \times 10^{44} \, \mathrm{erg \, s^{-1}}$, $R_\mathrm{cor} = 15 \, r_g$, $\tau_\mathrm{T} = 0.5$, $\xi_p = 0.1$, $\ell_c = 4.0 \, r_g$, $v_\mathrm{adv} = 0.03 \, c$, and $v_\mathrm{A} = 0.25 \, c$. These parameters are consistent  with previous studies \citep{dasRevealingProductionMechanism2024,lemoineNeutrinosStochasticAcceleration2025,Yuan2026}. We modelled the stochastic acceleration of protons using the diffusive acceleration scheme (Eq.~\ref{diffusive}). Alternative acceleration prescriptions are discussed in Appendix~\ref{sec:app}.

The resulting multi-messenger spectra are plotted in Fig.~\ref{fig:proton_photon_neutrino_stat}, which shows that the observed IceCube neutrino flux is satisfactorily reproduced by this standard set of parameters and that it remains consistent with the $\gamma$-ray constraints. This is a notable result that supports previous semi-analytical estimates. Note that Fig.~\ref{fig:proton_photon_neutrino_stat} includes two different IceCube reconstructions of the neutrino signal \citep[see][]{abbasiseyfert2025,abbasiEvidenceNeutrinoEmission2025}. The first (pink) corresponds to a phenomenological fit assuming a simple power-law spectrum, while the second (purple) is obtained under the assumption of a disk-corona emission model presented by \citet{muraseHiddenCoresActive2020,2021ApJ...922...45K}.
Since the disk-corona model spectrum provides a more physically motivated comparison with our model than a generic power-law, we focus on the latter (purple) dataset. We aim to show that a standard corona parametrisation produces neutrino emission comparable to IceCube reconstructed spectra, such that a direct comparison with the IceCube error bands is therefore sufficient, and we do not perform a likelihood fit.

The stationary proton spectrum (inset of Fig.~\ref{fig:proton_photon_neutrino_stat}) is relatively flat and the integrated energy density lies close to the background plasma pressure. Both features arise from the self-regulation induced by the backreaction of accelerated protons on the turbulent cascade. A cut-off appears at a few tens of TeV, corresponding to the energy at which photo-hadronic losses become dominant over acceleration (see Fig.~\ref{fig:timescales}). The associated neutrino spectrum (red curve in Fig.~\ref{fig:proton_photon_neutrino_stat}) reflects the underlying proton distribution. The $pp$ component is approximately flat and dominates at low energies ($\sim 1$–$100$ GeV). At higher energies ($\sim 1$–$30$ TeV), neutrino production is dominated by photo-hadronic ($p\gamma$) interactions, highlighting the crucial role of the coronal X-ray photon field, as noted in earlier studies. Protons at tens of TeV efficiently produce neutrinos in the $3$–$30$ TeV range observed by IceCube, making the $p\gamma$ threshold a natural explanation for the characteristic neutrino energies.

Figure~\ref{fig:proton_photon_neutrino_stat} also shows that \texttt{Turb-AM3} reproduces the main features of the coronal X-ray spectrum, arising from inverse Compton scattering of disk photons by thermal electrons. For typical coronal conditions \citep{rybickiRadiativeProcessesAstrophysics1979}, the X-ray luminosity spectrum expected from thermal Comptonization can be approximated as
\begin{equation}
L_{\rm X,\varepsilon} \propto \varepsilon^{2}\frac{dN_\gamma}{d\varepsilon} 
\propto \varepsilon^{(\ln \tau_T / \ln A)+1} 
\exp\!\left(-\frac{\varepsilon}{k_B T_e}\right),
\end{equation}
where $\tau_\mathrm{T}$ is the Thomson optical depth and 
$A = 1 + 4k_B T_e / (m_e c^2)$ is the average energy amplification factor per scattering. For $\tau_\mathrm{T} \lesssim 1$ and $A \gtrsim 1$, this expression yields an approximately flat spectrum 
extending up to an exponential cut-off at $\varepsilon \sim k_B T_e \sim 100\,\mathrm{keV}$.
Our results are consistent with this expectation: disk photons are efficiently up-scattered, producing a hard power-law X-ray spectrum with a high-energy cut-off at a few hundred keV. We note that \texttt{AM3} calculates a time-dependent photon distribution through repeated inverse Compton calculations, and that this distribution rapidly reaches the above equilibrium spectrum. In the current version, the electron distribution around the thermal peak is not finely resolved by the numerical grid. While the current implementation provides a satisfactory description of the X-ray emission, a more accurate treatment of thermal Comptonization will be addressed in future work. 

High-energy $\gamma$ rays produced in hadronic interactions initiate an electromagnetic cascade in the dense photon field of the corona. Absorption through $\gamma\gamma \rightarrow e^+e^-$ pair production generates secondary pairs that cool via synchrotron and inverse Compton emission, producing additional photons and sustaining the cascade. As a result, the initial TeV $\gamma$-ray power is redistributed to lower energies, suppressing the direct escape of TeV photons. This reprocessing is visible in Fig.~\ref{fig:proton_photon_neutrino_stat}, where a broad cascade extends from $\sim 1\,\mathrm{MeV}$ to $\sim 1\,\mathrm{GeV}$. The pair density produced by the cascade is $n_\mathrm{pairs} \simeq 5.0 \times 10^{6} \, \mathrm{cm^{-3}}$, much smaller than the primary electron density $n_{e^-} \simeq 2.2 \times 10^{10} \, \mathrm{cm^{-3}}$. Pair production therefore remains dynamically negligible and does not significantly modify the electron distribution, validating the assumption $n_{e^\pm} \approx n_{e^-}$ throughout the evolution. The evolution of the electron energy spectrum, including pair production, is shown in Appendix \ref{sec:elec_evol}.

Accordingly, the predicted high-energy gamma-ray spectrum lies below the observed level of emission, whose origin is attributed to the extended starburst region \citep[e.g.][]{ajelloDisentanglingHadronicComponents2023a}. At lower energies ($\lesssim1$ MeV), disk and coronal photons are absorbed and reprocessed by the dusty torus. Within these constraints, our model reproduces the IceCube neutrino flux without violating $\gamma$-ray limits, consistent with \citet{dasRevealingProductionMechanism2024}.

To ensure the physical viability of the model, we performed a comprehensive consistency check of the energy budget. The total coronal + disk photon luminosity is $L_\mathrm{\gamma,tot} = 7.6 \times 10^{44} \, \mathrm{erg \, s^{-1}}$, which remains below the Eddington limit ($L_\mathrm{edd} \simeq 2.1 \times 10^{45} \, \mathrm{erg \, s^{-1}}$). 
The Eddington ratio is therefore $\lambda_\mathrm{Edd} \equiv L_\mathrm{\gamma,tot}/L_\mathrm{edd} \simeq 0.36$, consistent with \citet{dasRevealingProductionMechanism2024}.
The total photon luminosity is also in good agreement with the bolometric luminosity reported by \citet{Woo_2002}.
Furthermore, our modelled $(2-10)$ keV luminosity ($L_\mathrm{X} \simeq 6.8 \times 10^{43} \, \mathrm{erg \, s^{-1}}$) aligns with observational constraints \citep{marinucciNuSTARCatchesUnveiling2016}.
The energy budget is also partitioned into a magnetic and a proton component. We can define the non-thermal proton luminosity with the accretion luminosity or with the diffusive escape luminosity, so $L_p \simeq 4 \pi R_\mathrm{cor}^2v\int\varepsilon_p (dN_p/d\varepsilon_p) d\varepsilon_p$ with $v = R_\mathrm{cor}/t_\mathrm{adv}$ or  $v = R_\mathrm{cor}/t_\mathrm{esc}$ respectively. This results in $L_p \simeq 2.6 \times 10^{43} \, \mathrm{erg \, s^{-1}}$ or $L_p \simeq 4.8 \times 10^{43} \, \mathrm{erg \, s^{-1}}$ respectively. 
Either way, the derived proton power is consistent with previous estimates \citep{dasRevealingProductionMechanism2024} and remains safely below the bolometric luminosity. 
The accreting turbulent luminosity is $L_{\delta B} = 1.2 \times 10^{43} \, \mathrm{erg \, s^{-1}}$.
Together, these results indicate that the proposed scenario is energetically robust.
However, the solution is not unique, as illustrated by the parameter degeneracies shown in Figs.~\ref{fig:stasol} and \ref{fig:esc_acc_cst}.

\begin{figure}
\resizebox{\hsize}{!}{\includegraphics[clip=true]{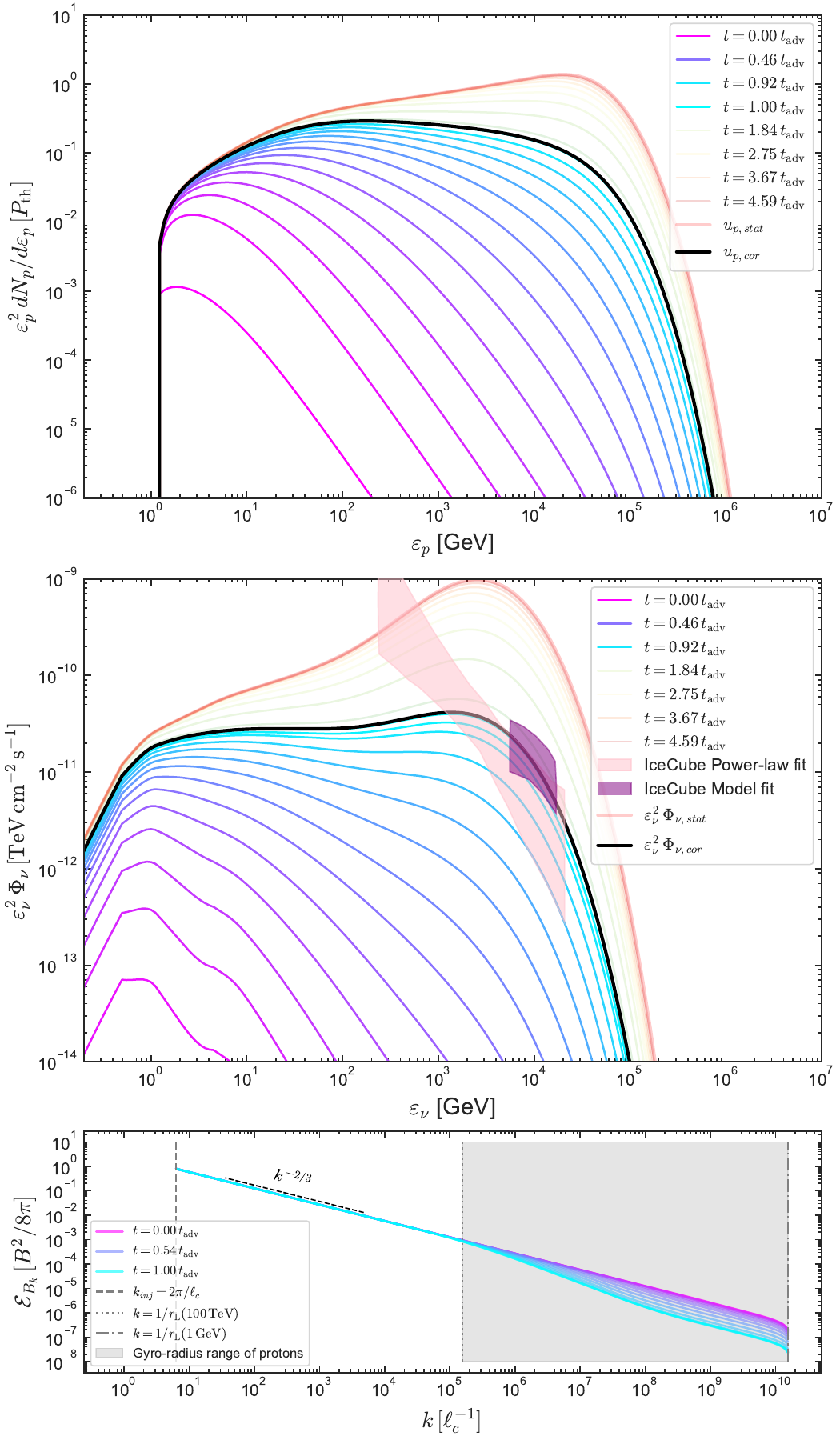}}
\caption{
\textit{Top}: Time evolution of the integrated proton SED over the advection time $t_\mathrm{adv}$ 
(purple to blue), i.e. the plasma crossing time of the corona; the spectrum is expressed in units of the background plasma pressure. 
The stationary proton SED in the corona is shown in black. The green–orange curves display the proton SED evolution beyond 
$t_\mathrm{adv}$. The red curve corresponds to the proton SED obtained by integrating up to times well in excess of $t_\mathrm{adv}$ 
(for illustration purposes). These spectra correspond to the steady state solutions of Eq.~(\ref{eq:transport}). \textit{Middle}: Same as the 
top panel but for the neutrino flux. \textit{Bottom}: Time evolution of the magnetic turbulent cascade during the advection time. The 
grey region highlights the part of the inertial range that dominates particle acceleration of protons up to $100\,$TeV.
}
\label{fig:proton_photon_neutrino_evol}
\end{figure}

The spectrum shown in Fig.~\ref{fig:proton_photon_neutrino_stat} corresponds to our set of fiducial parameters. We examine its sensitivity to this choice of parameters in the following section. Here, we discuss how the overall spectral shape departs from other models proposed in the literature. First, this spectrum represents an average over a time-dependent evolution through the coronal volume, while other models generically evaluate the spectrum as the steady state solution to the transport equation. To better visualise this time evolution, we plot in Fig.~\ref{fig:proton_photon_neutrino_evol} the time-dependent spectra of a proton population (top panel) injected at an initial time $t=0$ together with the corresponding secondary neutrino spectra (middle panel). In those panels, we also show the average spectrum representative of a stationary corona, obtained on timescales $t\gg t_{\rm adv}, t_{\rm esc}$. Early on $(t\lesssim 0.2t_{\rm adv}$), the proton spectrum $\varepsilon_p^2 {\rm d}N_p/{\rm d}\varepsilon_p$ grows in intensity and to larger energies due to acceleration. This is in agreement with the standard result ${\rm d}N_p/{\rm d}\varepsilon_p\propto \varepsilon_p^{-1-t_{\rm acc}/t_{\rm esc}}$ in the absence of energy losses and backreaction on the turbulence~\citep{1954ApJ...119....1F}, implying that for $t_{\rm acc}<t_{\rm esc}$, as is the case here (Fig.~\ref{fig:timescales}), $\varepsilon_p^2 {\rm d}N_p/{\rm d}\varepsilon_p$ is an increasing function of $\varepsilon_p$. However, once the proton energy density comes close to the turbulent energy density, damping becomes effective and non-linear backreaction becomes effective. This effect is visible in the plot of the turbulent energy power spectrum in the lower panel, which reveals damping. In agreement with \citet{lemoineNonlinearAspectsStochastic2024}, the acceleration rate drops for low energy particles, while higher-energy particles keep gaining energy by interacting with larger-scale modes. The proton spectrum then levels out, shaping a proton spectrum with, approximately, ${\rm d}N_p/{\rm d}\varepsilon_p \propto \varepsilon_p^{-2}$. A clear signature of this effect is to produce a flat extension to the neutrino spectrum, below the peak energy at $\sim\,$TeV energies, that results from $pp$ interactions. Models that do not include the backreaction of accelerated particles on the turbulence, or that assume $\xi_p\lesssim 10^{-3}$ (see thereafter), rather produce a neutrino spectral shape with a pronounced peak at TeV energies.

In our framework, the physically relevant stationary spectrum for the whole corona is set by $t=t_\mathrm{adv}$, since advection limits particle residence time in the corona. If acceleration could continue past $t_{\rm adv}$, the spectrum would eventually reach a steady state regime corresponding to $\partial_t \mathcal{N}_p \approx 0$ in Eq.~(\ref{eq:transport}), shown by the green to orange curves in Fig.~\ref{fig:proton_photon_neutrino_evol}. This occurs at times exceeding the diffusive escape and energy loss timescales, i.e. at $t \gtrsim 5\,t_\mathrm{adv}$ (red curve).
As discussed in the following section, such a steady state could be achieved within one $t_{\rm adv}$ with a faster acceleration rate, for example a higher $v_{\rm A}$. This would produce a mild pile-up in the proton spectrum, as a result of the competition between acceleration and energy losses. This pile-up would  be pronounced in the neutrino spectrum as a result of the strong dependence of $p\gamma$ losses on proton energy (see Fig.~\ref{fig:proton_photon_neutrino_evol}). The overall neutrino spectral shape would then display a characteristic pronounced peak at TeV energies, similar to those obtained in the literature not accounting for the feedback on turbulence \citep[e.g.][]{Yuan2026}.

Such models can account for the observed neutrino spectrum at the price of tuning the proton injection fraction $\xi_p$, or the filling fraction of active regions in the corona, by the right amount. The dependency of the spectral shape on this  parameter is now examined.

\subsection{Exploration of parameter space}\label{sec:scan}
For given ambient physical conditions, the rates of energy loss are fixed, and hence the proton spectral shape is mostly governed by three parameters: the injection fraction ($\xi_p$) and the ratios of advection or escape to acceleration timescales (Sect.~\ref{sec:transport}). We explore the influence of these various parameters in the present section to assess the sensitivity of the predicted neutrino spectral shapes on the choice of parameters. 

\begin{figure}
\includegraphics[width=\columnwidth]{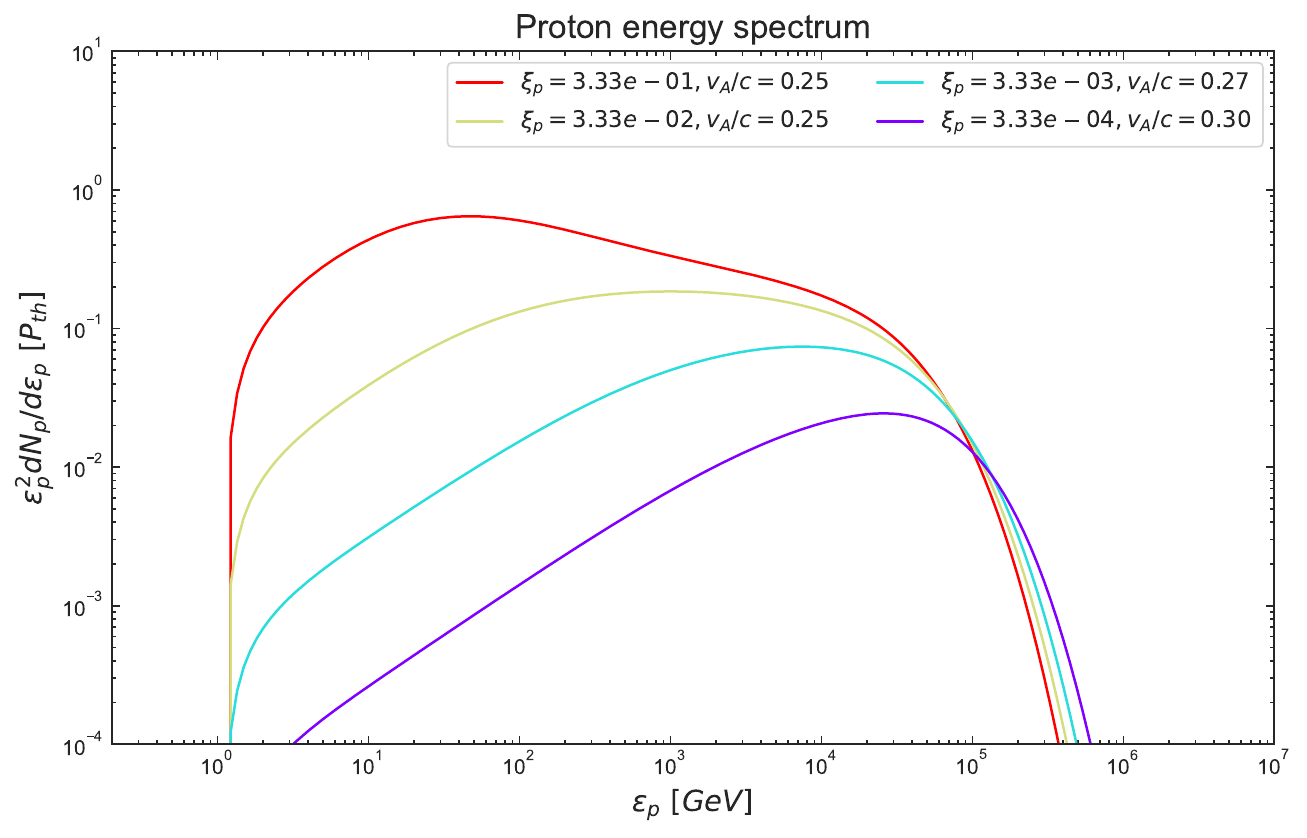}
\includegraphics[width=\columnwidth]{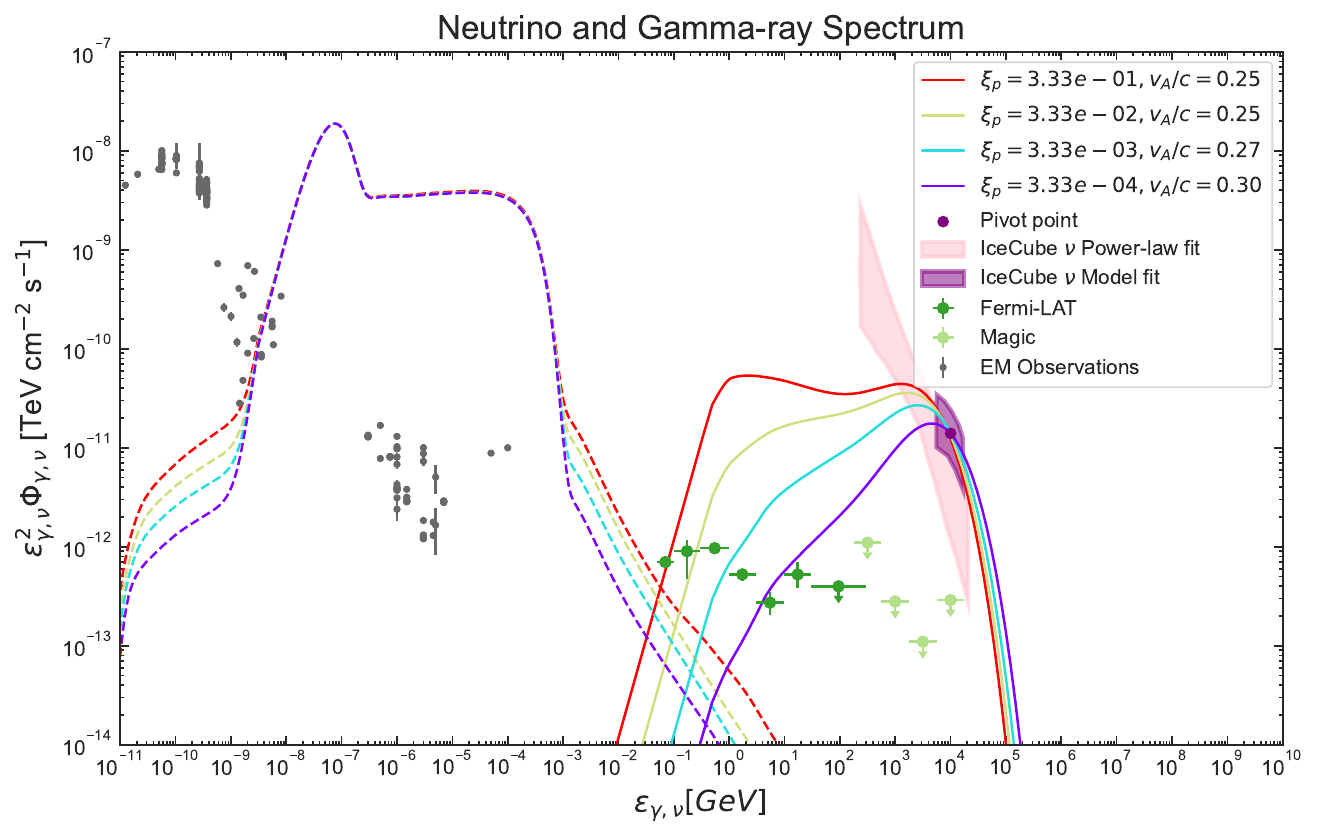}
\caption{ Integrated proton (\textit{top}) as well as photon and neutrino stationary (\textit{bottom}) energy spectra. The initial energy fraction in non-thermal protons and the Alfvénic velocity are indicated for each model. The other parameters are $L_d = 5.0 \times 10^{44} \, \mathrm{erg \, s^{-1}}$, $R_\mathrm{cor} = 15 \, r_g$, $\ell_c = 4.0 \, r_g$, and $v_\mathrm{adv} = 0.03 \, c$.}
\label{fig:stasol}
\end{figure}

Figure~\ref{fig:stasol} shows the spectra for different non-thermal energy fractions, $\xi_p$. To obtain a meaningful comparison to observations, we slightly tuned the Alfvénic velocity ($v_{\rm A}$)  to match the observed neutrino flux at a pivot point defined by $E_\nu = 10\,$TeV and (all-flavour) flux $E_\nu^2 \Phi_\nu = 1.4\times 10^{-11}\,$erg/cm$^2$s (see the corresponding figure). In practice, this implies increasing $v_{\rm A}$ as one decreases $\xi_p$; other parameters remain fixed. One key observation is that, at low values of $\xi_p$, the neutrino spectrum becomes increasingly peaked. This occurs because turbulent damping is weak when $\xi_p< 10^{-2}$, implying that proton acceleration mostly takes place in the linear (or test-particle) regime. As noted earlier, the proton spectral shape then becomes hard, with approximately $\varepsilon_p^2{\rm d}N_p/{\rm d}\varepsilon_p\propto \varepsilon_p^{1-t_{\rm acc}/t_{\rm esc}}$. In turn, this shapes a steeper $pp$ contribution to the neutrino flux.

At moderate values of the injection fraction and above, $\xi_p\gtrsim 10^{-2}$, proton acceleration becomes self-regulated by turbulent damping, and the proton spectrum converges towards a universal saturation state, whose total energy density is of the order of the total turbulent energy content. The  model reliably reproduces the IceCube neutrino flux,
independently of the precise value of $\xi_p$, shaping a neutrino spectral shape with a turn-over at energies below the peak ($E_\nu \sim 3\,$TeV here). The flat extension to lower energies is prominent for the largest value of $\xi_p$. As mentioned above, it is shaped by the $pp$ contribution that follows the approximately flat proton spectrum. 

We remark here that this contribution would be suppressed in proportion to $n_p/n_e$ if pair loading becomes substantial (here, $n_p/n_e=1$). This scaling arises from the observation that the ratio $n_p/n_e$ governs the plasma proton density, with $n_e$ determined by the opacity constraint, yet it has little effect on the high-energy non-thermal proton content. The overall pressure of these protons indeed remains comparable to the turbulent magnetic pressure once turbulent damping becomes significant (see Eq.~\ref{eq:Pnth-selfr}). Future high-sensitivity neutrino observations of NGC~1068 in this energy range could help discriminate between these  possibilities. 

\begin{figure}
\includegraphics[width=\columnwidth]{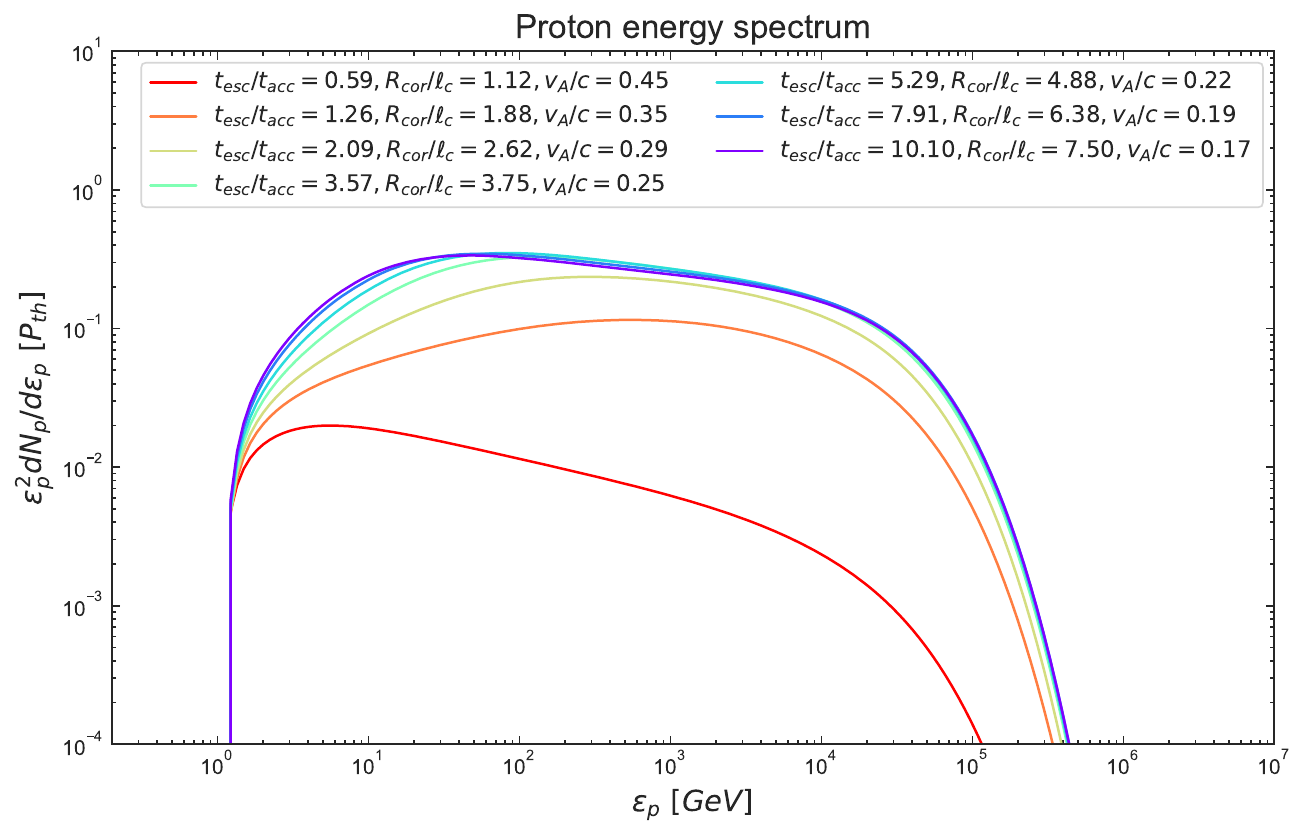}
\includegraphics[width=\columnwidth]{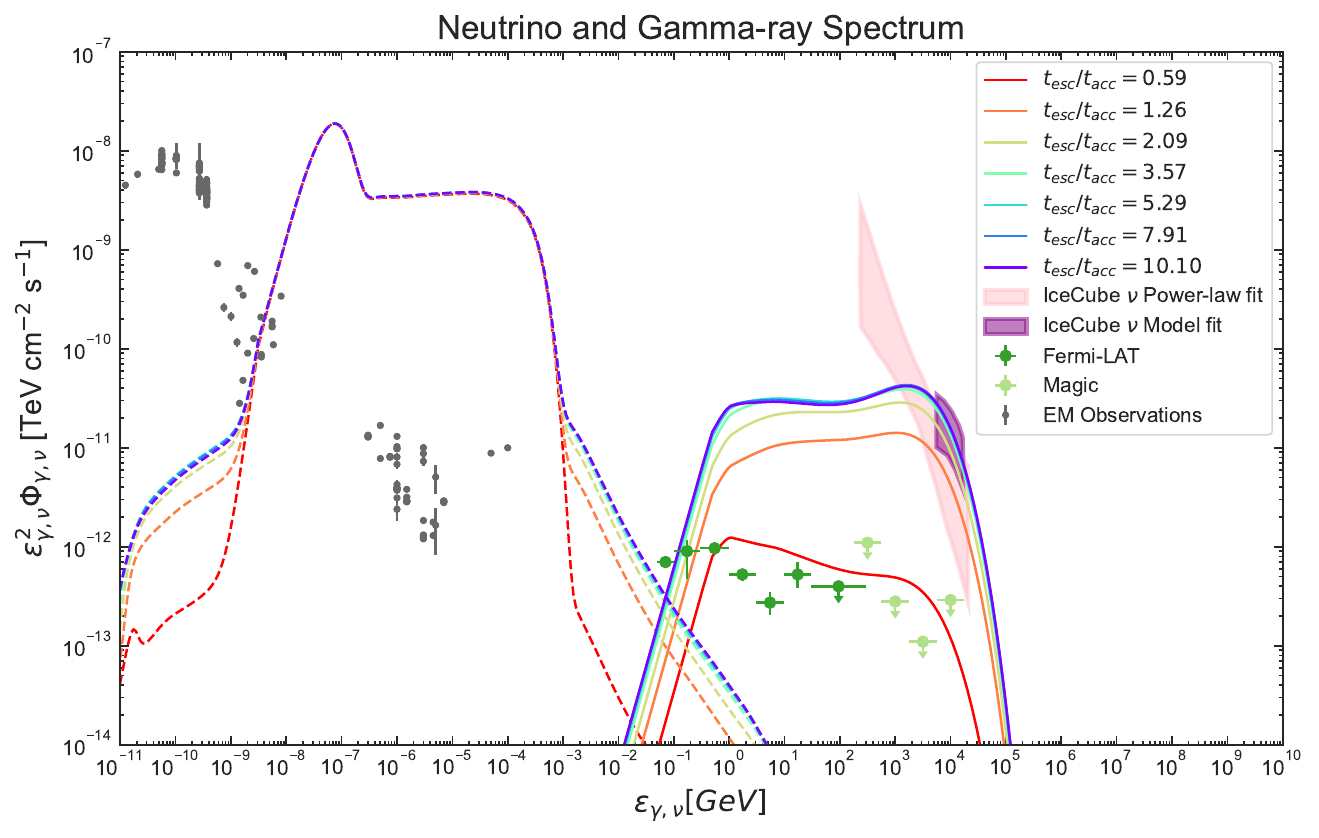}
\caption{Proton, photon, and neutrino stationary spectra for different coronal microphysics parameters, characterised by varying dimensionless time ratios, $t_{\rm esc}/t_{\rm acc}$. The parameters $v_\mathrm{A}$ and $\ell_c$ were varied, while the others were fixed to $L_d = 5.0 \times 10^{44} \, \mathrm{erg\,s^{-1}}$, $R_\mathrm{cor} = 15 \, r_g$, $\xi_p = 0.1$, and $v_\mathrm{adv} = 0.03 \, c$. The same colour in both panels corresponds to the same parameter set.}
\label{fig:esc_acc_cst}
\end{figure}

We now turn to the influence of the ratio between the proton escape and acceleration timescales, $t_{\rm esc}/t_{\rm acc}$ (Eq.~\ref{eq:def-Xi}), while maintaining a constant ratio between the advection and acceleration timescales, whose influence is discussed later. We satisfied these requirements by varying $\ell_{\rm c}$, hence the ratio $R_{\rm cor}/\ell_{\rm c}$ at constant $R_{\rm cor}=15\,r_{\rm g}$, together with $v_{\rm A}\propto (R_{\rm cor}/\ell_{\rm c})^{-1/2}$. Figure~\ref{fig:esc_acc_cst} illustrates the characteristic dependence of the various spectra on this ratio $t_{\rm esc}/t_{\rm acc}$. It confirms the anticipated result that $t_{\rm esc}/ t_{\rm acc}\gtrsim 1$ is a prerequisite to proton acceleration to high energies. In this limit, the proton spectrum acquires a near universal shape with approximately equal energy per decade, as discussed in the non-thermal energy fraction study. This is a direct consequence of turbulence damping, which becomes more prevalent when protons are strongly confined, as one would anticipate. 
As $t_{\rm esc}/t_{\rm acc}\gtrsim 1$ takes larger values, the normalisation of the proton spectrum rises due to more efficient confinement. However, this growth is ultimately limited by the finite turbulent energy reservoir, leading to a saturation of the proton energy density at large values of $t_{\rm esc}/t_{\rm acc}$.
The corresponding neutrino spectra also exhibit the universal shape when the turbulence feedback is significant and naturally reproduce the observed neutrino spectrum for $t_{\rm esc}/t_{\rm acc} > 1$. 

\begin{figure}
\includegraphics[width=\columnwidth]{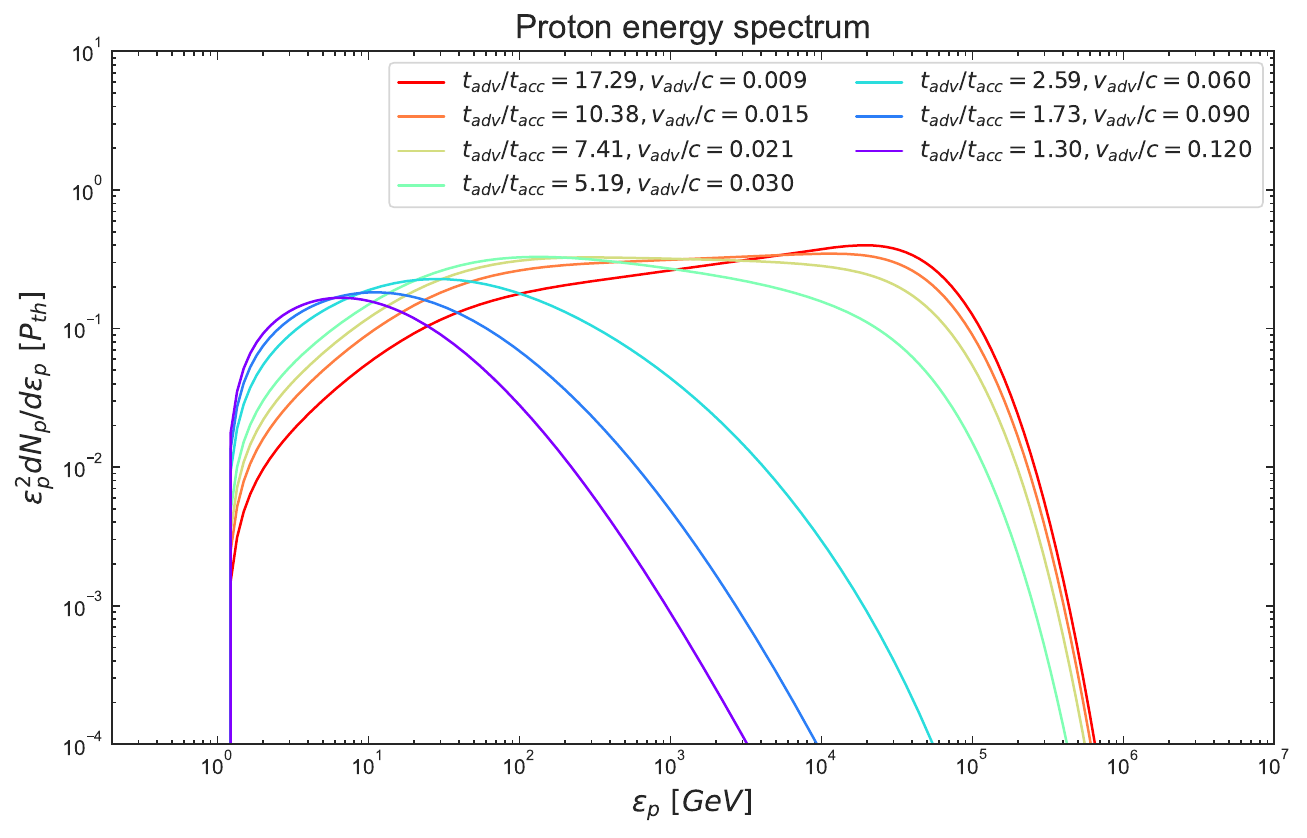}
\includegraphics[width=\columnwidth]{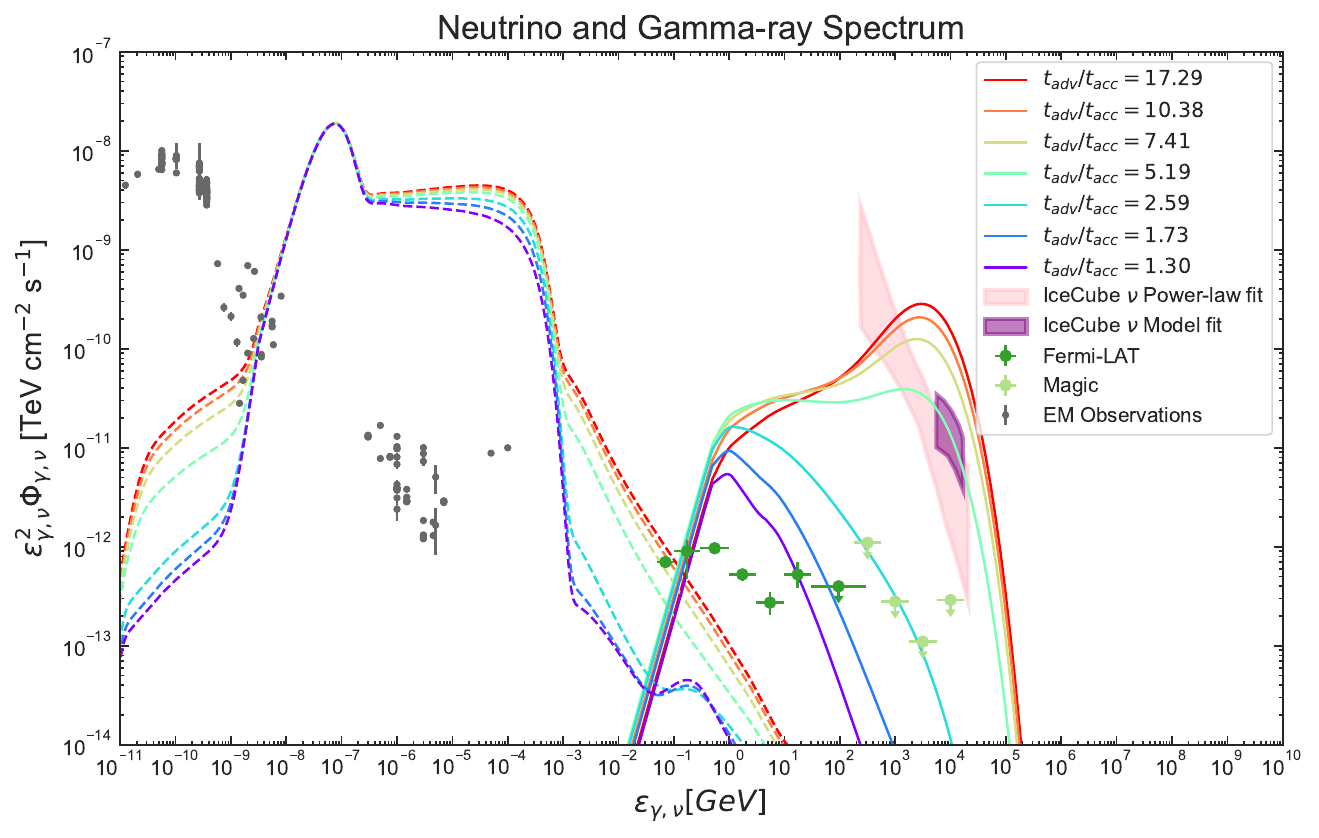}
\caption{Same as Fig.~\ref{fig:esc_acc_cst} but for different time ratios $t_{\rm adv}/t_{\rm acc}$. Further, $v_\mathrm{adv}$ was varied. The other parameters are $L_d = 5.0 \times 
10^{44} \, \mathrm{erg \, s^{-1}}$, $R_\mathrm{cor} = 15 \, r_g$, $\xi_p = 0.1$, $v_\mathrm{A} = 0.25 \, c$, and $\ell_c=4.0\, r_g$. }
\label{fig:adv_acc}
\end{figure}

To gauge the influence of varying $t_{\rm adv}/t_{\rm acc}$ while maintaining $t_{\rm esc}/t_{\rm acc}$ fixed, we varied $v_{\rm adv}$ with all other parameters fixed to their value in Sect.~\ref{sec:fiducial}. We recall that the value chosen so far, $v_{\rm adv}=0.03\,c$, corresponds to the standard radial inflow velocity at a position $R_{\rm cor}=15\,r_{\rm g}$. Varying this parameter thus offers a way of exploring different corona scenarios just as it tests our assumption of spatially uniform $v_{\rm adv}$. As expected, increasing $t_{\rm adv}/t_{\rm acc}$ shifts the energy cut-off towards higher values (see Fig.~\ref{fig:adv_acc}).
This is because the advection timescale $t_\mathrm{adv}$ sets the duration over which protons are accelerated and therefore determines the maximum achievable energy. In the diffusive regime, neglecting escape and radiative losses, the mean proton momentum evolves as $\langle p \rangle_{t = t_\mathrm{adv}} = \langle p \rangle_{t = 0}\,\exp\!\left(4 t_{\rm adv}/t_{\rm acc}\right)$. Thus, for the stationary proton population to reach the characteristic energies $E \sim 10-100$~TeV required to reproduce the observed neutrino flux, $t_{\rm adv}/t_{\rm acc}$ must be of the order of a few.
However, for large values of $t_{\rm adv}/t_{\rm acc}$, the cut-off energies converge between $10$ and $100$~TeV. This behaviour arises because, once protons reach the energy at which photo-hadronic ($p\gamma$) losses dominate over acceleration, the losses timescale, rather than the advection timescale, determine the cut-off.
In the limit of large $t_{\rm adv}/t_{\rm acc}$, the stationary spectrum asymptotically approaches the solution obtained when advective proton losses are neglected, corresponding to the red curve in Fig.~\ref{fig:proton_photon_neutrino_evol}. 
For $v_\mathrm{adv} < 0.02\,c$ the multi-messenger flux exceeds the IceCube bound and the $\gamma$-ray constraints, unless $\xi_p$ is artificially reduced to low values. A clear measurement of the neutrino spectrum in the corresponding range could potentially disentangle the various possibilities and thus shed light on the inner physics of the corona.

Finally, we briefly address the influence of the microphysical transport model on the neutrino spectral shape. Our parameter study has thus far relied on the standard diffusive Fokker-Planck framework, with parameters derived from numerical kinetic simulations. However, exploring the alternative models outlined in Sect.~\ref{sec:acc} revealed that the neutrino spectral shape remains relatively unaffected by this choice in the self-regulated regime. This insensitivity arises because the proton spectrum adopts a nearly universal flat shape under these conditions. This is demonstrated in Fig.~\ref{fig:acc_schemes}. 

\subsection{A template neutrino average spectrum}\label{sec:universal}
This exploration of parameter space reveals that the proton and neutrino  energy spectra take different shapes depending on the ratios $t_{\rm adv}/t_{\rm acc}$ and $t_{\rm esc}/t_{\rm acc}$. The injected proton fraction $\xi_p$ does not strongly influence the spectral shape provided $\xi_p\gtrsim 10^{-2}$. In a situation such as the present one, where the spectral shape varies substantially as a function of one or two parameters whose actual values cannot take precisely uniform values through the corona, or precisely constant ones in time, the actual spectral shape is bound to represent an average of the various possibilities. We thus explore here the possibility of an extended distribution of these parameters. This extended distribution can be understood in various ways. For instance, the physical conditions can vary on timescales larger than $t_{\rm adv}$, implying that the observed neutrino spectrum is effectively a time average. Or, the corona itself can comprise patchy regions of small extent that each contribute to energisation with local conditions. 

\begin{figure}[h]
\includegraphics[width=\columnwidth]{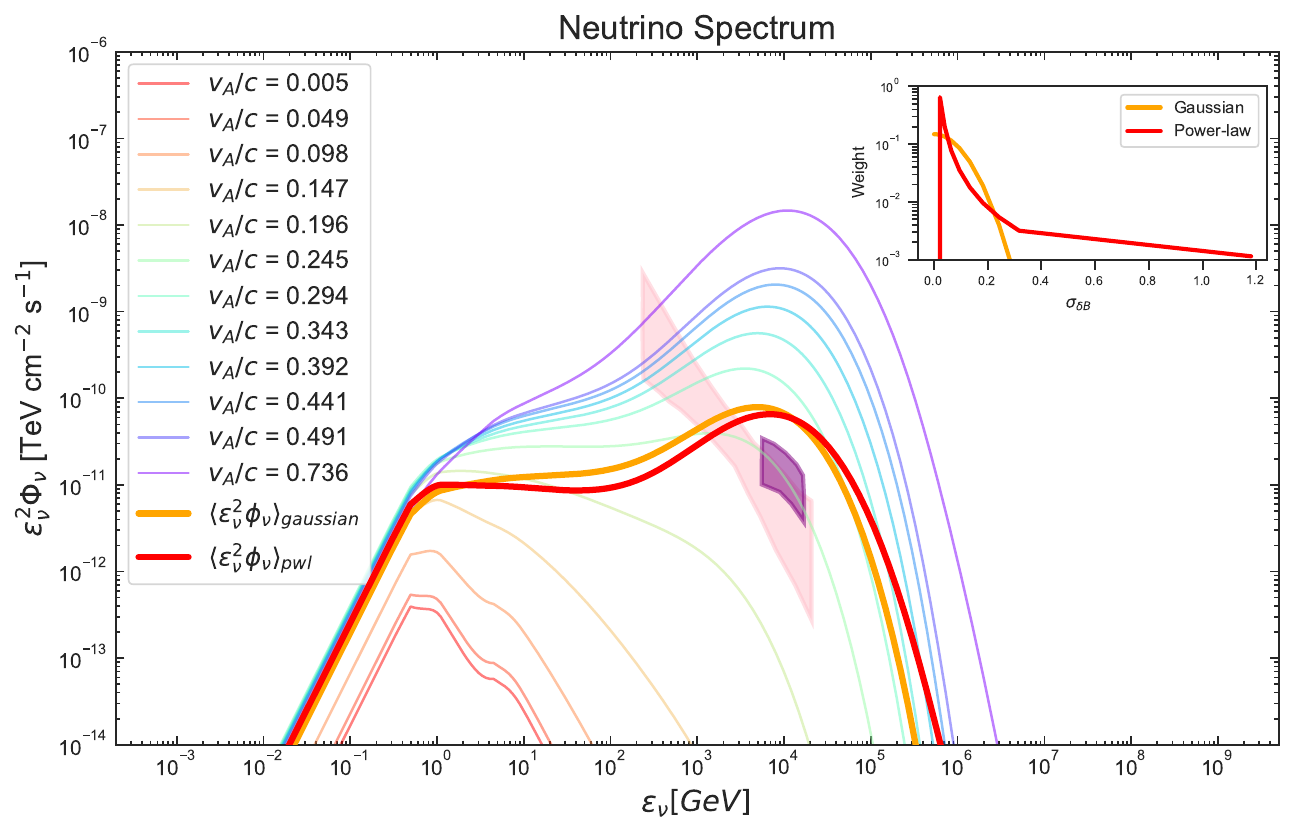}
\caption{Neutrino spectra obtained by averaging over different realisations of the coronal magnetisation, $\sigma_{\delta B}$, treated as a random 
variable controlling the acceleration rate. Two weighting schemes are shown: a Gaussian distribution and a power-law distribution $w(\sigma)\propto 
\sigma^{-2}$, both normalised to the same mean magnetisation $\langle \sigma \rangle = 0.04$.}
\label{fig:weighted-nu}
\end{figure}

We simplified the problem by assuming that the magnetisation $\sigma_{\delta B}$, which controls the acceleration rate (Eq.~\ref{eq:tacc}), is the only parameter subject to spatial fluctuations. We modelled it as a random variable with a prescribed distribution $w(\sigma)$, normalised such that the mean value $\langle \sigma \rangle = \int \mathrm{d}\sigma\,\sigma\,w(\sigma)$ is fixed to $0.04$ (corresponding to $v_\mathrm{A} \simeq 0.2\,c$), consistent with the coronal conditions discussed in Sect.~\ref{coronamodel}. For illustration, we considered both a Gaussian distribution and a power-law distribution of weights, $w(\sigma) \propto \sigma^{-k}$ with $k=2$, over the interval $\sigma \in [0.01,\,2]$ \citep[see e.g.][]{groseljHighenergyEmissionTurbulent2026}. 
In practice we computed neutrino spectra as in Sect.~\ref{sec:fiducial}, but for different magnetisations. We then averaged the neutrino spectra according to the prescription in $w(\sigma)$.
The resulting (averaged) neutrino spectra are shown in Fig.~\ref{fig:weighted-nu}. We adopted a non-thermal proton fraction $\xi_p = 0.1$, in line with our fiducial value, for which the non-thermal proton energy density takes values of the order of the turbulent energy density and does not strongly affect the spectral shape.
We find that the average spectral shape is insensitive to the choice of weight distribution. The resulting neutrino spectrum exhibits the characteristic `universal' profile: a broad peak with a low-energy extension dominated by $pp$ interactions, tracing the underlying proton density. As discussed above, this component could be suppressed in the presence of significant pair loading in the corona. At higher energies, the spectrum extends beyond the peak compared to the fiducial single-zone model (Fig.~\ref{fig:proton_photon_neutrino_stat}), reflecting the contribution of intermittent (in space or time) regions with locally enhanced acceleration efficiency. 
Nevertheless, the averaged spectrum still displays a pronounced cut-off at a few tens of TeV, set by photo-hadronic losses. This indicates that, within the present framework, it remains challenging to account for neutrino emission extending beyond $\sim 100$~TeV. 

\subsection{Extension to other Seyfert galaxies}
\label{otherseyferts}

\begin{figure}
    \includegraphics[width=\columnwidth]{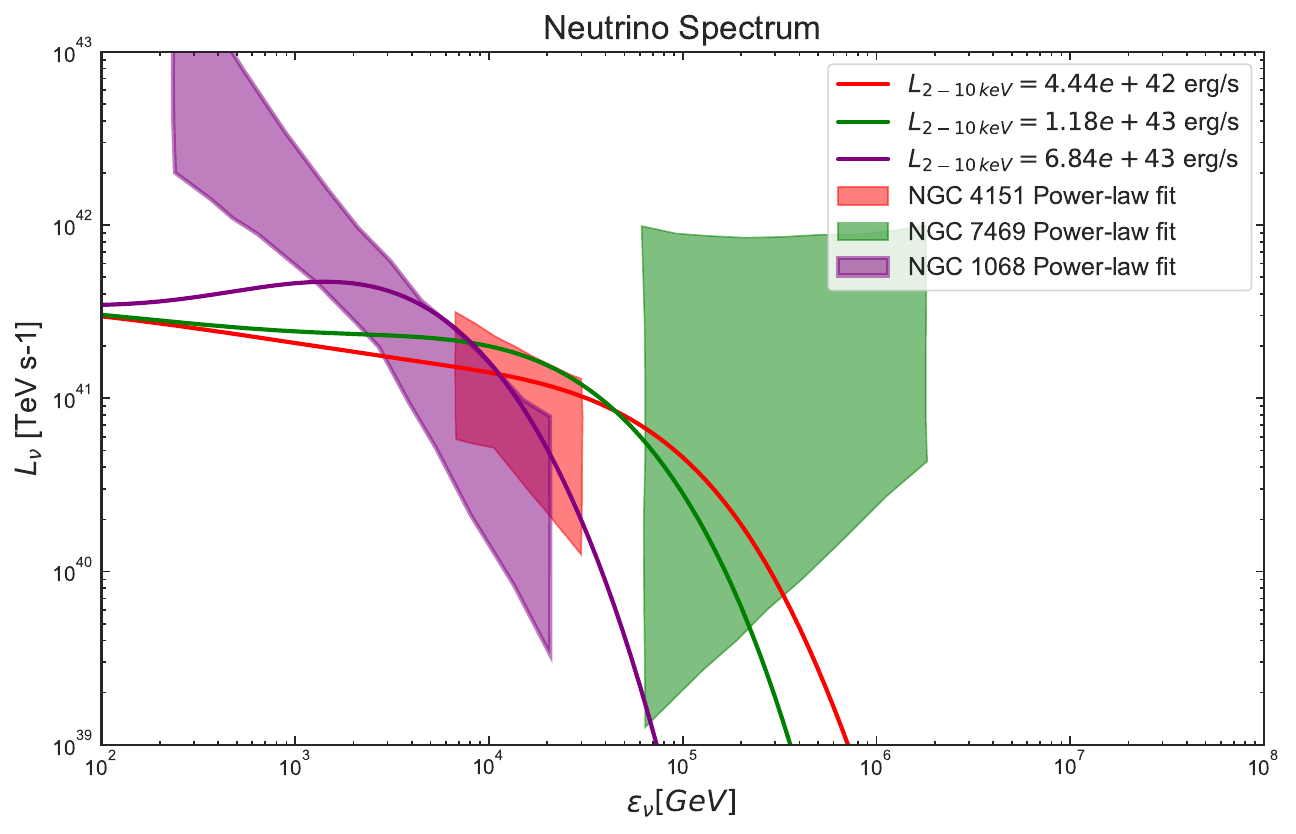}
    \includegraphics[width=\columnwidth]{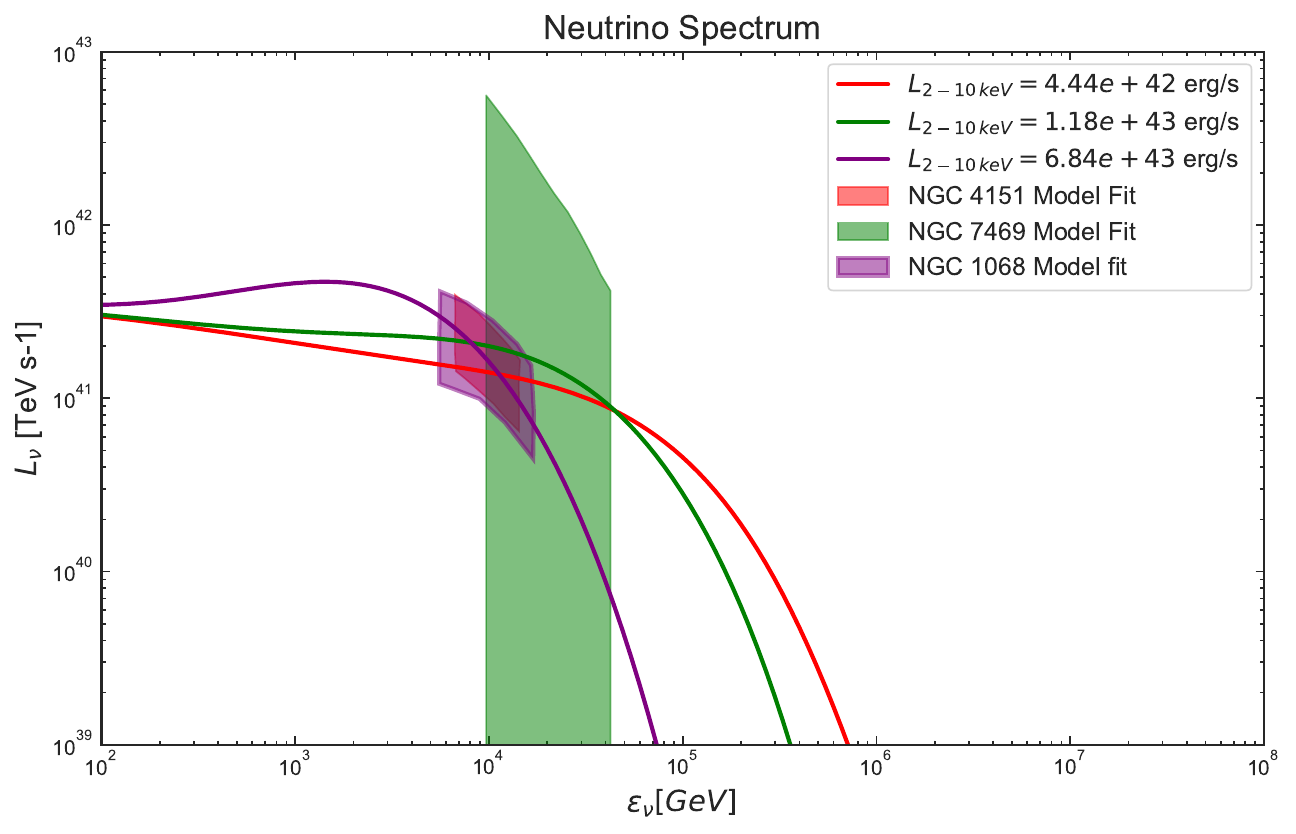}
    \caption{Predicted stationary neutrino spectra for different Seyfert galaxies compared with IceCube inferred fluxes. 
    \textit{Top:} Comparison with IceCube results obtained under a power-law assumption. 
    \textit{Bottom:} Comparison with IceCube results inferred using a disk–corona model. 
    In both panels, model predictions were computed using the same fiducial parameters as in Sect.~\ref{sec:fiducial}, with variations in the disk luminosity.}
    \label{fig:otherplots}
\end{figure}

A natural extension of this work is to explore how our model applies to other X-ray–bright Seyfert galaxies.
Recent IceCube observations \citep{abbasiEvidenceNeutrinoEmission2025} suggest a possible trend in the $(F_\nu, E_\nu)$ plane: sources detected at higher characteristic neutrino energies tend to exhibit lower fluxes (see Fig.~\ref{fig:otherplots}), with the notable exception of CGCG~420$-$015. Such a trend could potentially result from a varying X-ray luminosity~\citep{2026arXiv260220969Y}.
An increase in coronal luminosity enhances the density of target photons (if the size of the corona does not vary), thereby boosting the efficiency of $p\gamma$ interactions. This leads to a higher neutrino flux, but also to stronger photo-hadronic cooling, which shifts the proton cut-off, and consequently the neutrino cut-off, to lower energies. Conversely, lower luminosities reduce the neutrino flux while allowing protons to reach higher maximum energies, shifting the neutrino emission towards higher energies. This mechanism can therefore account for part of the observed diversity in the neutrino properties of Seyfert galaxies.

Figure~\ref{fig:otherplots} compares the neutrino luminosities inferred by IceCube for NGC~1068, NGC~4151, and NGC~7469 (under a power-law assumption or a disk-corona model assumption), with the predictions of our model. Starting from the fiducial parameter set described in Sect.~\ref{sec:fiducial}, we vary only the disk luminosity (and thus the coronal X-ray luminosity) to match the observed $2-10$~keV luminosities of NGC~4151 \citep[$L_{2-10\,\mathrm{keV}} = 5^{+3}_{-2} \times 10^{42}\,\mathrm{erg\,s^{-1}}$;][]{kumarNGC4151} and NGC~7469 \citep[$L_{2-10\,\mathrm{keV}} = (1.0-1.7) \times 10^{43}\,\mathrm{erg\,s^{-1}}$;][]{prince2025echomappingblackhole}. 
Despite its simplicity, this approach reproduces the neutrino emission levels of these Seyfert galaxies reasonably well. In particular, it accommodates, within the relatively large error bars, the potential $\sim 100$~TeV neutrino detection that could be associated with NGC~7469, highlighting the ability of the model to capture the broad range of observed neutrino energies.

\section{Conclusions}\label{sec:conc}
We have presented a self-consistent numerical framework to model high-energy neutrino production in radiatively dense environments, and we applied it to the case of NGC~1068 to derive state-of-the-art predictions for the neutrino spectrum in the TeV range. This numerical framework leverages the implementation of hadronic and radiative loss channels and the self-consistent treatment of electromagnetic cascades in the numerical tool \texttt{AM3}~\citep{klingerAM$^3$OpenSourceTool2024}. With respect to \texttt{AM3}, a new and noteworthy feature is to offer a self-consistent time-dependent model of stochastic acceleration for the protons while properly including the backreaction of proton acceleration on the turbulent cascade (and hence on the acceleration itself) and incorporating detailed prescriptions from microphysics. One advantage of this approach is that it provides a unified description of particle acceleration and multi-messenger emission in environments that are possibly radiatively dense, as is the case for hidden neutrino sources. In our work, we kept the electron distribution function fixed, as the electrons evolve on short timescales compared to those of protons in the case investigated. However, future studies should incorporate detailed energy loss and energisation processes for mildly relativistic electrons, as in existing models of X-ray emission from black hole coronae, to improve self-consistency.

We used the introduced numerical framework to examine the case of NGC~1068 as a high-energy neutrino source in Sect.~\ref{sec:results}. We have demonstrated that the observed neutrino spectrum of NGC~1068 can be satisfactorily reproduced using physically motivated standard parameters. In particular, a coronal size of $R_\mathrm{cor} \sim 5-20\,r_g$, an Alfvén velocity of $v_\mathrm{A} = \mathcal{O}(0.2\,c)$, a coherence length of $\ell_c \sim R_\mathrm{cor}/4$, and an advection velocity of $v_\mathrm{adv} = \mathcal{O}(0.03\,c)$ yield acceleration, escape, and advection timescales satisfying $t_\mathrm{acc} \lesssim t_\mathrm{esc}, t_\mathrm{adv}$, allowing protons to reach the required energies. The inferred magnetic field strength, $\delta B = \mathcal{O}(10^3)$\,G, assuming near-equipartition, and a proton-to-lepton ratio of $n_p/n_e \lesssim 1$ are consistent with standard coronal models. The resulting neutrino emission is compatible with current $\gamma$-ray constraints and in agreement with previous studies \citep{eichmannSolvingMultimessengerPuzzle2022,fangHighenergyNeutrinosInner2023,dasRevealingProductionMechanism2024,Yuan2026}.  

For the given set of parameters, the coupled evolution of proton acceleration and turbulence establishes a natural self-regulating mechanism. The accelerated protons, through turbulent damping, extract energy from the cascade, thereby curbing their own growth and maintaining the non-thermal proton energy density at a level comparable to that of the turbulent energy density. This self-regulation provides a natural explanation for the normalisation of the neutrino flux observed from NGC~1068.

We have also demonstrated that this model successfully reproduces the IceCube data, within the relatively large error bars, for the Seyfert galaxies NGC~4151 and NGC~7469. We used identical parameters except for their X-ray luminosity, which was adjusted to match observed values.

Our exploration of the parameter space also allowed us to extract a generic template for the neutrino spectrum: a gradual rise to a peak around $\sim 1-10\,$TeV followed by a decline with an approximately power-law-like behaviour over a limited energy range. This high-energy suppression directly reflects the cut-off in the proton distribution due to efficient photo-hadronic losses. Notably, the nearly flat extension below the peak, shaped by $pp$ interactions, serves as a useful indicator of the proton density in the corona and, consequently, the pair-loading factor. 

Overall, the interplay between magnetic turbulence and radiative processes in the corona sets both the normalisation and the shape of the neutrino emission. In this sense, neutrino observations already provide a direct probe of coronal plasma conditions despite existing parameter degeneracies.

\begin{acknowledgements}
The authors are grateful to D.~Groselj for insightful discussions. This work was supported by the French
Agence Nationale de la Recherche, ANR, project ANR-25-CE31-3279.
\end{acknowledgements}

\bibliographystyle{aa_url.bst}
\bibliography{Article_NGC1068_2026_SLB_ML_FR}

\begin{appendix}
\section{Proton, photon, and neutrino spectra for different acceleration schemes}
\label{sec:app}

In this appendix we present a comparative exploration of the impact of different acceleration schemes on the 
resulting particle spectra.

\begin{figure}[h]

    \includegraphics[width=\columnwidth]{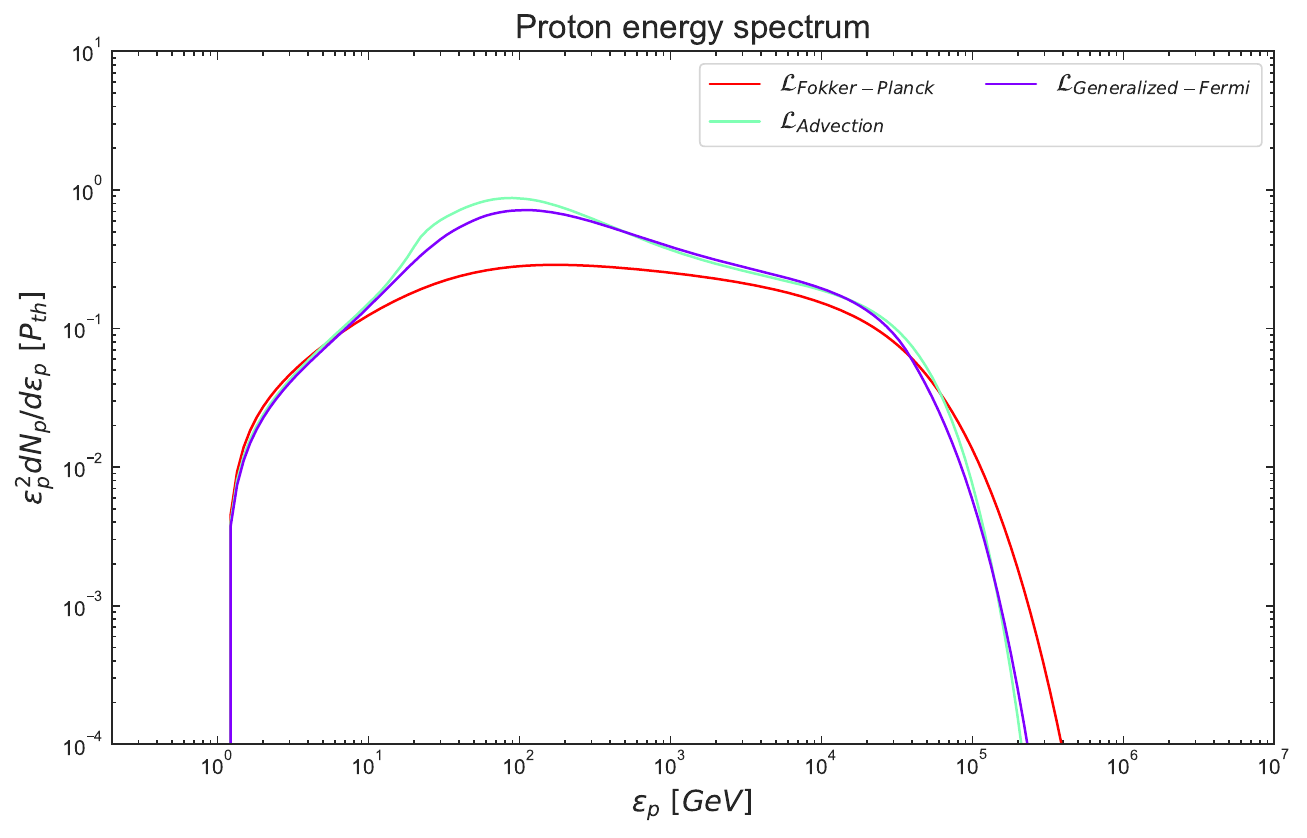}

    \includegraphics[width=\columnwidth]{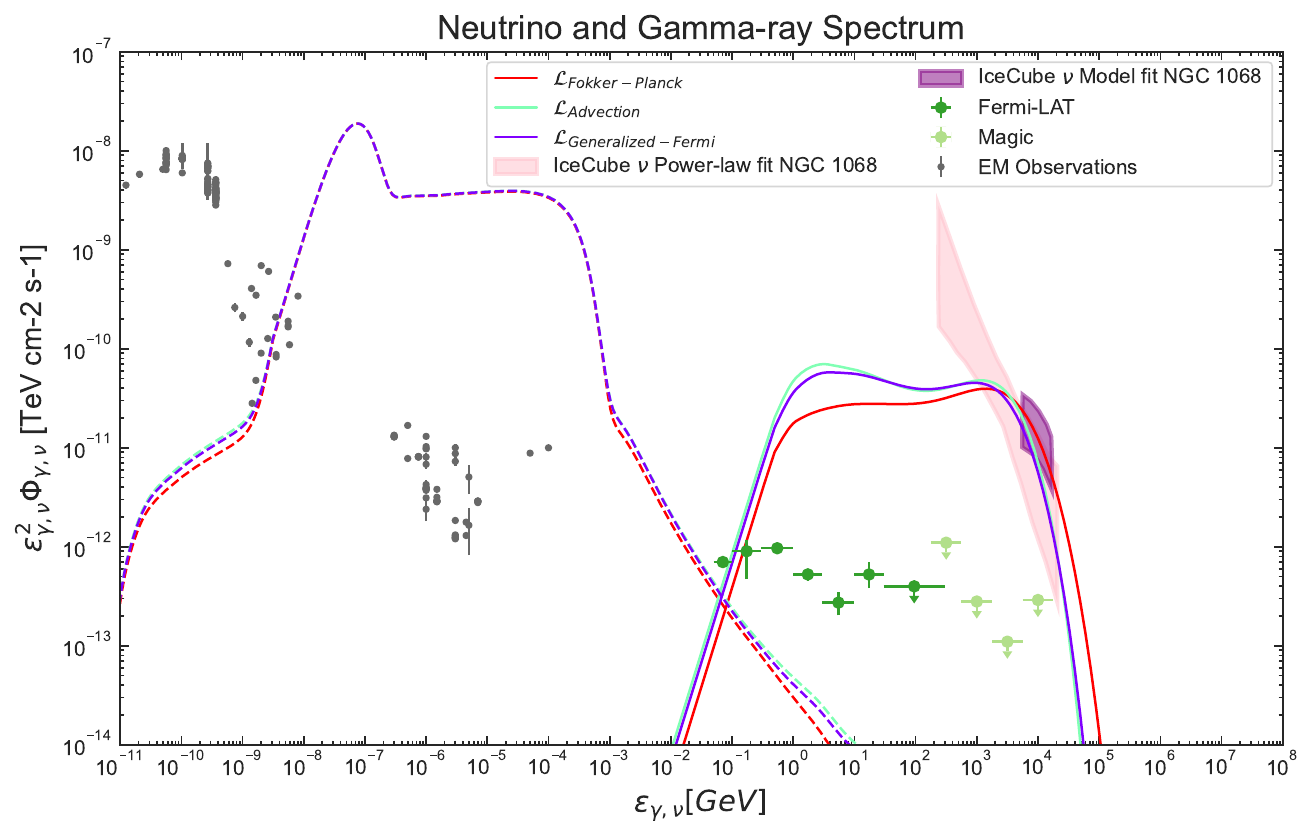}

    \caption{Proton, photon, and neutrino stationary spectra with different microphysics description of the proton acceleration. For each acceleration scheme we use our fiducial parameter set (see Sect. \ref{sec:fiducial}).}
    \label{fig:acc_schemes}
\end{figure}

In Fig.~\ref{fig:acc_schemes}, we compare several prescriptions used to model stochastic particle acceleration 
(see Sect.~\ref{sec:acc}): the Fokker–Planck (diffusive) scheme, the generalised Fermi approach, and a first-order 
(`Fermi I') scheme. In the latter case, stochastic acceleration is replaced by a systematic drift in momentum 
space. 
This contribution is described by an advective term in momentum space, 
\begin{equation}
    \mathcal{L}_{\rm Adv} \mathcal{N}_p \equiv -\,\partial_p\!\left(A_p\, \mathcal{N}_p\right),
    \label{advective}
\end{equation}
where $A_p$ is an effective momentum drift coefficient. The corresponding acceleration rate is defined as 
$\nu_{\mathrm{acc}} \equiv A_p/p$. Although the exact momentum dependence of $A_p$ is uncertain, we adopted here a 
constant acceleration rate,
\[
\nu_{\mathrm{acc}} \simeq \frac{4 v_\mathrm{A}^2}{3\,\ell_c\,c},
\]
chosen such that the systematic acceleration operates with an efficiency comparable to that of diffusive stochastic 
acceleration.

These schemes are evaluated for the same set of physical parameters in order to isolate the role of the microphysics. We 
find that their impact on the resulting proton and neutrino spectra remains limited, provided that the injected 
non-thermal proton energy fraction is sufficiently large. In this regime, the system enters a feedback-dominated state 
in which the proton population self-regulates through its interaction with the turbulent cascade.

As a result, the various acceleration prescriptions lead to qualitatively similar spectra: the proton spectral energy density 
saturates at a level comparable to the thermal plasma pressure, and exhibits a high-energy cut-off at a few tens of TeV 
due to efficient photo-hadronic losses. The associated neutrino spectra inherit these features.

\section{Evolution of the electron spectrum}
\label{sec:elec_evol}
\begin{figure}[h]
\includegraphics[width=\columnwidth]{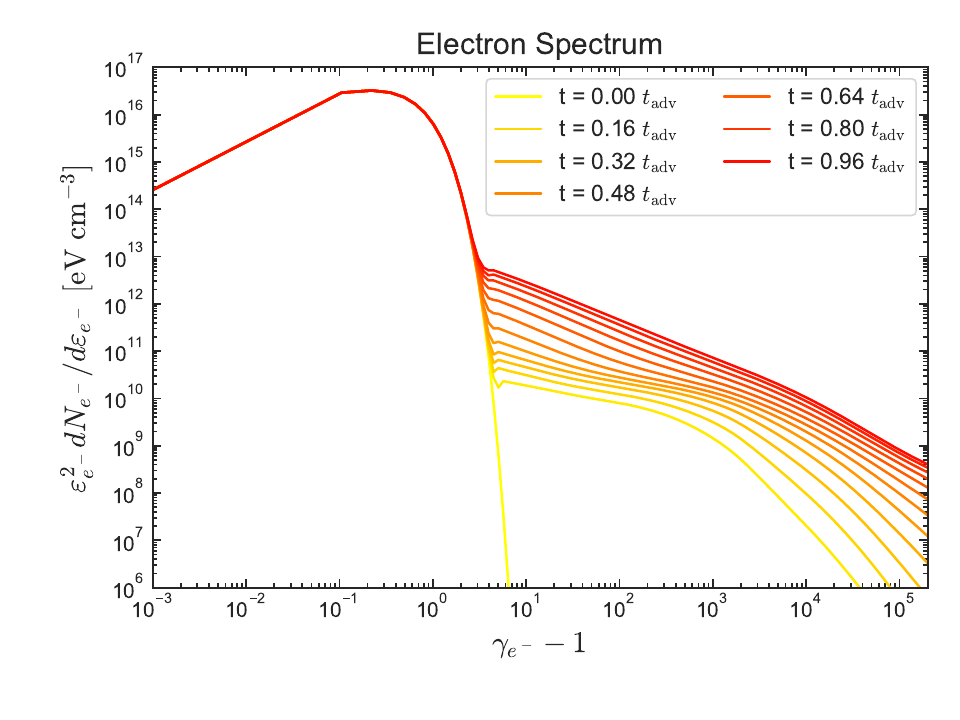}
\caption{Time evolution (yellow to red) of the electron energy spectrum up to $t_{\rm adv}$, with our fiducial parameter set (see Sect.\ref{sec:fiducial}).}
\label{fig:electron_spectrum}
\end{figure}

Figure~\ref{fig:electron_spectrum} shows the time evolution of the electron energy distribution $\varepsilon_{e^-}^2 dN_{e^-}/d\varepsilon_{e^-}$. Two components are visible: a constant in time thermal Maxwell-J\"uttner distribution at low energies, responsible for the Comptonization of disk photons up to $\varepsilon_{\rm cut} \sim k_BT_e \sim 100\,\mathrm{keV}$, and a growing non-thermal tail at $\gamma_e \gtrsim$ a few, fed by secondary pairs from hadronic processes and $\gamma\gamma$ interactions, which contributes to the high-energy photon emission.

\FloatBarrier 
\twocolumn

\end{appendix}
\end{document}